\documentclass[a4paper,superscriptaddress,twocolumn,prd,showkeys]{revtex4}
\newcommand{\HENus}{high-energy neutrinos }
\newcommand{\HENu}{high-energy neutrino }

\newcommand{\figref}{Fig.~\ref}

\usepackage{graphicx,amsmath,subfigure,lipsum}
\usepackage{longtable}
\usepackage{color}

\begin{document}

\title{Obscured flat spectrum radio AGN as sources of \HENus} 

\author{G. Maggi}
\email{giuliano.maggi.olmedo@gmail.com}
\affiliation{Vrije Universiteit Brussel, Dienst ELEM, IIHE, Pleinlaan 2, 1050 Brussels, Belgium}

\author{S. Buitink}
\affiliation{Vrije Universiteit Brussel, Department of Physics and Astrophysics, Pleinlaan 2, 1050 Brussels, Belgium}

\author{P. Correa}
\affiliation{Vrije Universiteit Brussel, Dienst ELEM, IIHE, Pleinlaan 2, 1050 Brussels, Belgium}

\author{K.D. de Vries}
\email{krijndevries@gmail.com}
\affiliation{Vrije Universiteit Brussel, Dienst ELEM, IIHE, Pleinlaan 2, 1050 Brussels, Belgium}

\author{G. Gentile}
\affiliation{Sterrenkundig Observatorium, Universiteit Gent, Krijgslaan 281, 9000 Gent, Belgium}
\affiliation{Vrije Universiteit Brussel, Department of Physics and Astrophysics, Pleinlaan 2, 1050 Brussels, Belgium}

\author{J. Le\'on Tavares}
\affiliation{Sterrenkundig Observatorium, Universiteit Gent, Krijgslaan 281, 9000 Gent, Belgium}
\affiliation{Instituto Nacional de Astrof\'isica \'Optica y Electr\'onica (INAOE), Apartado Postal 51 y 216 72000, Puebla, Mexico}

\author{O. Scholten}
\affiliation{Vrije Universiteit Brussel, Dienst ELEM, IIHE, Pleinlaan 2, 1050 Brussels, Belgium}
\affiliation{University of Groningen, KVI-Center for Advanced Radiation Technology, Groningen, The Netherlands}

\author{N. van Eijndhoven}
\email{nickve.nl@gmail.com}
\affiliation{Vrije Universiteit Brussel, Dienst ELEM, IIHE, Pleinlaan 2, 1050 Brussels, Belgium}

\author{M. Vereecken}
\affiliation{Vrije Universiteit Brussel, Dienst ELEM, IIHE and International Solvay Institutes, Pleinlaan 2, 1050 Brussels, Belgium}

\author{T. Winchen}
\affiliation{Vrije Universiteit Brussel, Department of Physics and Astrophysics, Pleinlaan 2, 1050 Brussels, Belgium}

\keywords{neutrino astronomy, radio galaxy, AGN obscuration, jet-matter interaction}

\begin{abstract}
Active Galactic Nuclei (AGN) are believed to be one of the main source candidates for the high-energy (TeV-PeV) cosmic neutrino flux recently discovered by the IceCube neutrino observatory. Nevertheless, several correlation studies between AGN and the cosmic neutrinos detected by IceCube show no significance. Therefore, in this article we consider a specific sub-class of AGN for which an increased neutrino production is expected. This sub-class contains AGN for which their high-energy jet is pointing toward Earth. Furthermore, we impose the condition that the jet is obscured by gas or dust surrounding the AGN.

A method is presented to determine the total column density of the obscuring medium, which is probed by determining the relative X-ray attenuation with respect to the radio flux as obtained from the AGN spectrum. The total column density allows us to probe the interaction of the jet with the surrounding matter which leads to additional neutrino production. Finally, starting from two different source catalogs, this method is applied to specify a sample of low redshift radio galaxies for which an increased neutrino production is expected.
\end{abstract}

\maketitle

\section{Introduction}\label{uhecr}
The Ultra-High-Energy Cosmic Ray (UHECR) spectrum has been mapped out by the Pierre Auger Observatory~\cite{augerres} and the Telescope Array~\cite{tares} collaborations. Nevertheless, the origin of these particles is still unknown. In case of a cosmic accelerator, the produced UHECRs can interact with ambient photons in the acceleration region producing high-energy neutrinos~\cite{mannheim,waxman, guetta, grb_meszaro, waxman_nu_astronomy, agn_stecker,grb_rachen_meszaro, ahlers, murase14,alv-muniz,halzen_zas,mucke03}. Since neutrinos are weakly interacting particles and have no charge, they are able to travel large distances through the universe without being deflected by the (inter)galactic magnetic fields. Therefore, if detected, these neutrinos point back to their production site, giving direct evidence for hadronic acceleration at their source.

The field of neutrino astronomy obtained a boost with the detection of the high-energy astrophysical neutrino flux by the IceCube neutrino observatory~\cite{i3_28, i3_37}. Nevertheless, even though several point-source searches were performed by the IceCube collaboration, no sources have been detected so-far~\cite{juanan,ps_i3_timedep,ic_6yr}. Different source candidates such as Gamma Ray Bursts (GRBs), Active Galactic Nuclei (AGN), star-forming galaxies, and supernova remnants (SNRs) have been investigated in detail by the IceCube collaboration, putting stringent limits on neutrino production models at these sites~\cite{juanan,ps_i3_timedep}. Furthermore, no clear anisotropy in the neutrino arrival directions has been observed~\cite{ic_6yr}.

One way to explain these results is to consider different neutrino production models in combination with a relatively high source density, leading to a more diffuse emission over the full sky. In this article we consider a different option by suggesting an up to now unexplored source class. We propose a sub-class of AGN as possible high-energy neutrino emitters. These sources are defined as obscured AGN with their high-energy jet pointing in the direction of Earth~\cite{our_icrc}. For this specific sub-set of AGN, the jet will be blocked by surrounding dust or gas giving rise to additional neutrino production through the jet-matter interaction in case a hadronic component is accelerated~\cite{murase2009}. It should be noted that not only the hadronic component inside the jet is blocked, but also the resulting gamma-ray flux. Hence the considered objects in this article compose a so-called hidden source population for GeV to TeV gamma-rays. Such a population is currently favored by IceCube and Fermi data interpreted in a multi-messenger approach~\cite{murase2016}.

We present a method to determine the total column density of the jet-obscuring matter, which allows us to estimate the additional neutrino production. The method is based on the X-ray attenuation as determined from the AGN spectrum. The X-ray attenuation typically reaches over a few orders of magnitude, which, as will be shown in section~\ref{proton_matter_subsec}, indicates that the obscuring dust or gas is located at parsec scales from the central engine. A possible explanation for such a component can be given by an orientation of the disc or dust torus covering the AGN jet, which has been considered in several AGN models~\cite{andy,bianchi,
risaliti_agnview,nature_titedtorus,elvis_obscuring_struct,risaliti_effect_bars}. 

As a first application of the developed method, we investigate two different galaxy catalogs to specify possible obscured flat spectrum radio AGN. The first catalog consists of a sample of galaxies with strong radio emission located in the close universe, from now on referred to as the Nijmegen catalog~\cite{nijmegen}. The second catalog consists of a sub-sample of the Fermi 2LAC catalog~\cite{fermi2lac}, containing mainly Flat Spectrum Radio Quasars (FSRQ) and BL Lac objects. First a red-shift selection is made to obtain an unbiased set of objects. Subsequently, to assure the high-energy jet is pointing toward us, a selection on the radio spectral index is made. Finally, we apply the method outlined in section~\ref{proton_matter_subsec} to specify possible obscured flat spectrum radio AGN. In case a hadronic component is accelerated inside the high-energy AGN-jet, these objects are expected to give rise to an increased neutrino production compared to non-obscured AGN.

\section{AGN as possible \HENu sources}\label{agn_section}
Active Galactic Nuclei (AGN) are considered to be the compact core of a galaxy with an extremely massive black hole in its center. Within the unified AGN model~\cite{agnbook}, the AGN consists of a sub-parsec scale disc of hot matter falling into the Black Hole, the so-called accretion disc. Further out, at parsec scales, the AGN is composed of a dusty torus surrounding the central Black Hole. Perpendicular to the plane of the accretion disc, many AGN show an outflow of highly relativistic particles which can extend up to several hundreds of parsecs or even kilo-parsec scales, the AGN-jet. It should be noted that where the accretion disc is generally considered perpendicular to the AGN-jet, several models exist where the dust torus can be tilted with respect to the accretion disc~\cite{andy,bianchi,risaliti_agnview,nature_titedtorus,elvis_obscuring_struct,risaliti_effect_bars}. 

The separate AGN components have been derived from the observed highly non-thermal AGN Spectral Energy Distributions (SEDs). The spectra are typically characterized by a two `bump' structure, where the first bump is believed to originate from the synchrotron emission of highly relativistic leptons in the AGN-jet. This emission is typically observed at frequencies in the radio band. The second bump can be shifted up to X-ray or even $\gamma$-ray energies. This bump originates from either low energy photons which are up-scattered to higher energies through the inverse Compton process or thermal emission from the hotter parts of the accretion disc. Another very interesting scenario would be the production of $\gamma$-rays through pion decay, directly indicating a hadronic component at these sources. Nevertheless, even though several models suggest the presence of some hadronic component, other models are able to explain the AGN spectra through electromagnetic interactions solely~\cite{leptonic_diltz,leptonic_hadronic_beammodel,leptonic_hadronic_reimer,leptonic_hadronic_m87}. 

Detailed studies of the AGN spectra allow to estimate several properties of the AGN such as their magnetic field, which is expected to be of the order of several Tesla up to parsec scales. Through the Hillas criterium, this immediately makes AGN one of the candidate sources for the origin of UHECRs, and hence high-energy cosmic neutrinos~\cite{hillas}.

\subsection*{Source Investigation}
The IceCube collaboration has investigated AGN as possible sources of \HENu emission. The latest results can be found in~\cite{juanan,ps_i3_timedep}. However, so-far no strong evidence for neutrino emission from AGN has been observed. The IceCube results presented in~\cite{juanan} are based on four years of IceCube data. A stacking analysis is performed on astrophysical objects for which a strong neutrino emission is expected through the photo-hadronic ($p\gamma$) neutrino production channel. The main feature of the selected objects is that they are very bright objects in the sky, with emission at high energies in the electromagnetic spectrum. Different types of astrophysical objects are investigated in this paper such as Local Starburst Galaxies, SNRs associated with molecular clouds, and AGN. In this search, no evidence for \HENu emission was obtained. A separate analysis has been developed in~\cite{ps_i3_timedep}, which is based on a Blazar population emitting over a broad range of the electromagnetic spectrum. Two different situations have been considered in that analysis, the first search considered neutrino emission during Blazar flares at random positions in the sky, and a second search has been performed looking for coincidences with $\gamma$-rays detected by Fermi-LAT~\cite{fermi_lat}. Again no significant neutrino emission from these source classes has been observed. Furthermore, the IceCube collaboration has put very stringent upper-limits on the neutrino production in GRBs~\cite{i3_grblimit}.

The natural question that arises is, what type of astrophysical objects are emitting the high-energy neutrinos detected by IceCube? Where many searches so-far considered objects which show strong emission at high energies in the electromagnetic spectrum, we propose in this article a source class for which this component is suppressed relative to the low-energy emission. The proposed cause for such a damping of the high-energy electromagnetic emission would be the attenuation due to surrounding gas or dust. Consequently, if a hadronic component is accelerated in the AGN-jet, this component will also be damped by the surrounding dust or gas. It follows that we expect a suppression in the output of UHECRs, where the jet-matter interaction will directly lead to additional high-energy neutrino production. Due to this additional neutrino production, the probability for a neutrino detector like IceCube to detect such a source increases, whereas the UHECR flux at Earth is expected to decrease.

\section{Neutrino production in AGN}\label{neutrino_emission}
The neutrino production mechanism in the high-energy jet of an AGN or GRB is expected to be similar in case a hadronic component is accelerated in their high-energy environment~\cite{halzen_zas,murase14}. In this situation, the high-energy hadronic component will interact with the ambient photon flux. In case of a pure proton composition accelerated at the source, the largest interaction cross-section is given by the production of the $\Delta$-resonance which gives rise to high-energy neutrinos through pion decay,
\begin{equation}\label{proton_gamma_pros}
p+\gamma \rightarrow \Delta^+ \rightarrow n+ \pi^{+} \rightarrow n + e^+ +\nu_e+ \nu_{\mu}+ \bar{\nu}_{\mu}.
\end{equation}

It should be noted that the neutron produced in~Eq.~\ref{proton_gamma_pros} has a relatively long lifetime, and at PeV energies is able to travel approximately 10~pc, escaping the containment region of protons before it decays into a proton and an electron accompanied by an anti-electron neutrino. Since the proton which originates from the neutron decay is outside of the containment region, it will be able to travel into outer space and might be responsible for the UHECR-flux measured at Earth~\cite{grb_rachen_meszaro,ahlers}. The anti-electron neutrino produced in the neutron decay has a much lower energy compared to the neutrinos produced by the decay of the $\pi^+$, and in case of a falling proton injection spectrum this component contributes negligibly to the overall high-energy neutrino spectrum. Several detailed models have been developed calculating the $p\gamma$ neutrino emission, where typically  5-20\% of the protons interact with the ambient photon flux~\cite{reimer2012}. In case composite nuclei are accelerated at the source, photodesintegration will take place by the ambient photon flux. As the energy to which a nucleus will be accelerated scales with its charge, the energy per nucleon is lowered by roughly a factor of two and the neutrino spectrum due to the $p\gamma$ interaction channel is shifted to lower energies.

\begin{figure}[!ht]
\centerline{
\includegraphics[width=.45\textwidth, keepaspectratio]{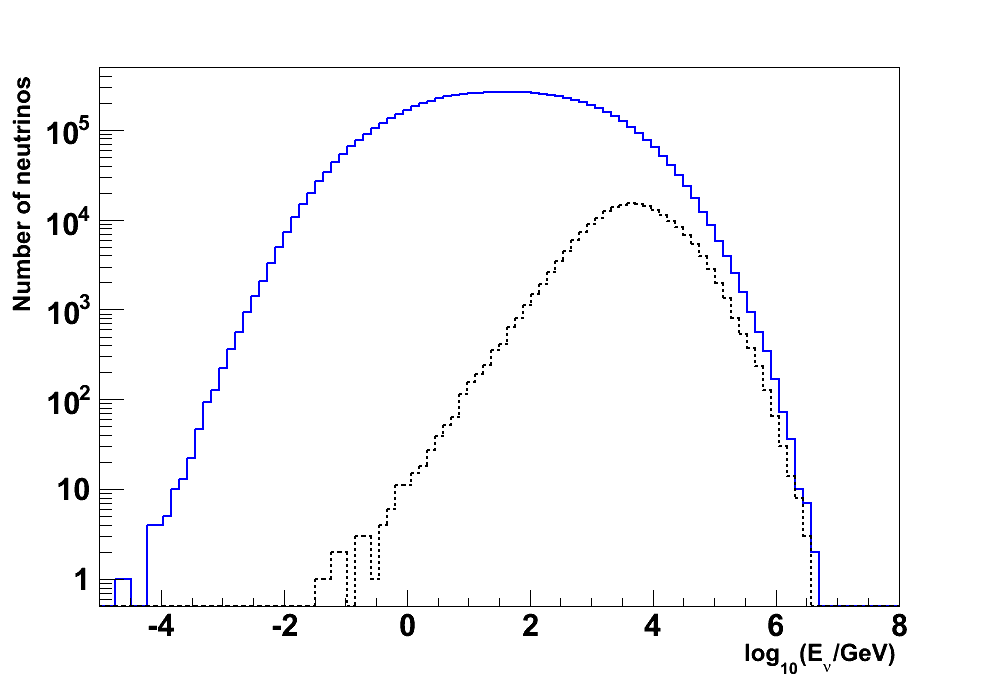}}
\caption{All flavor neutrino spectra for proton-proton $(pp)$ (blue solid line)
              and proton-gamma ($p\gamma$) (black dashed line) collisions. In total $10^5$ events were simulated following an $E^{-2}$ power-law spectrum between 0.2~PeV and 10~PeV. Of the generated events, 20000 (20\%) were forced to interact through the $p\gamma$ interaction channel, and 80000 (80\%) were forced to interact through the $pp$ interaction channel.}
\label{spectrum_fig}
\end{figure}

In addition to the $p\gamma$ induced neutrino flux we consider in this article a second contribution to the neutrino flux. This is through a jet obscuration by nearby matter, where the jet-matter interaction will lead to meson production which decay produces high-energy neutrinos. This production channel has been investigated using the PYTHIA 6.4 event generator~\cite{pythia}. A proton beam with a falling $E^{-2}$ energy spectrum between 0.2~PeV and 10~PeV has been simulated to interact head-on with photons of 10 keV energy in 20\% of the cases and with stationary protons in the remaining 80\% of the cases, thus reflecting a full ``beam dump" scenario. The total angle-integrated all flavor spectra are shown in~\figref{spectrum_fig}. It is seen that next to the neutrino production through the $p\gamma$ interaction channel (black dashed line), we expect a significant increase in neutrino production due to the proton-proton ($pp$) interaction channel (blue solid line). In addition to an increase of the peak flux, the neutrinos produced in the $pp$ interaction channel extend to lower energies compared to the $p\gamma$ interaction channel.

It follows that indeed a significant increase in the produced neutrino flux is expected through the jet-matter interaction in case the high-energy AGN jet is obscured by nearby dust or gas. The assumption that 20\% of the protons take part in $p\gamma$ interactions in the jet is generally believed to be an upper limit, where different proton energy loss processes such as synchrotron emission are neglected for the moment~\cite{reimer2012}. Furthermore, it should be noted that secondary baryons produced in the $p\gamma$ and $pp$ interactions will also interact with the surrounding dust. Including these processes in the spectrum calculation needs a more detailed dust or gas model, which goes beyond the scope of this paper and will be considered in a follow-up study. 

The increase of the neutrino flux for these objects will depend strongly on the fraction of protons which interact with the surrounding dust or gas. In addition it should be noted that in first approximation the neutrino production due to the jet-matter interaction and through the $p\gamma$ interaction are independent. To estimate the fraction of protons interacting in the dust, we discuss in the following section a method to determine the total column density of the obscuring dust or gas. Subsequently in section~\ref{blazar_selection}, we use the developed method to specify a set of possible obscured flat spectrum radio AGN.

Along with the production of additional high-energy neutrinos during the beam-dump process, in the obscured AGN scenario one also expects additional gamma-rays produced through neutral pion decay. In the unobscured AGN scenario, this process is lead by the counterpart of Eq.~1, $p\gamma \rightarrow \Delta^+ \rightarrow p + \pi^0$. Nevertheless, as will be shown in the following section, the typical thickness for a dust or gas cloud to have a significant increase in neutrino production through the $pp$ interaction channel equals several proton interaction lengths. At the considered energies, the proton interaction length is of the same order as the gamma-ray interaction length. Consequently, not only the protons will be absorbed by the dust or gas, but also the produced gamma rays. Therefore, even though some gamma rays might escape for deep interacting protons, in general we expect the gamma-ray flux to be highly suppressed for the objects considered in this article.

\section{Neutrino production in obscured flat spectrum radio AGN}\label{proton_matter_subsec}

\subsection{Jet-Matter interaction}
In this section we discuss the jet-matter interaction in more detail. A method is presented to determine the total amount of dust or gas from X-ray observations of the AGN. This allows us to determine the fraction of the high-energy hadronic component in the jet that will interact with the dust leading to additional neutrino production. A first application of this method will be presented in the following section to specify a set of obscured flat spectrum radio AGN.

The neutrino production due to the jet-matter interaction, $(p-N)$, scales with the relative amount of protons that will interact with the gas or dust,
\begin{equation}
\frac{I_p^{int}}{I_p^0}= 1-e^{-X_{tot}/\lambda_{p-N}},
\label{Ip}
\end{equation}
where $X_{tot}$ is the total column depth of the dust in which the jet-matter interaction occurs, $N$ denotes the dust constituent, and $\lambda_{p-N}$ is the proton interaction depth defined by the integrated density over the path length for which the number of protons drops with a factor $e^{-1}$. If $X_{tot}=4\lambda_{p-N}$, roughly 98\% of the protons interact, which gives rise to an effective beam dump process. The total column depth of the dust, $X_{tot}$, can be obtained through the X-ray attenuation. This is obtained by determining the X-ray intensity as measured at Earth, $I_X^{obs}$, relative to a base value $I_X^0$ in case there is no dust obscuration,
\begin{equation}
\frac{I_X^{obs}}{I_X^0}=e^{-X_{tot}/\lambda_{X}}.
\label{Ix}
\end{equation}
The determination of the base value, $I_X^0$, will be treated in the following section. Furthermore, we need to determine the X-ray attenuation depth in the dust, $\lambda_X$, as well as the proton interaction depth, $\lambda_{p-N}$. 

Since at the energies in which we are interested no accelerator data is available for proton-matter collisions, we base ourselves on the proton-air interaction depth determined by the Pierre Auger collaboration~\cite{auger_proton_air_xsec}. This measurement gives a proton-air interaction depth of $\lambda_{p-air}=56\;\mathrm{g\;cm^{-2}}$ for a mean proton energy of $E=10^{18.24}$~eV.

In view of the availability of measurements, we consider in this article an X-ray energy of 1.24 keV. To determine the X-ray attenuation depth, we use data from the XCOM photon cross-section database~\cite{XCOM}. In principle the X-ray energy observed at Earth will be different from the one at the source because of cosmological redshift effects. However, as will be shown in the following section, we will limit our sample to low redshifts ($z<0.17$) so that this effect will be small and can therefore be ignored. 

\begin{figure}[!ht]
\centerline{
\includegraphics[width=.45\textwidth, keepaspectratio]{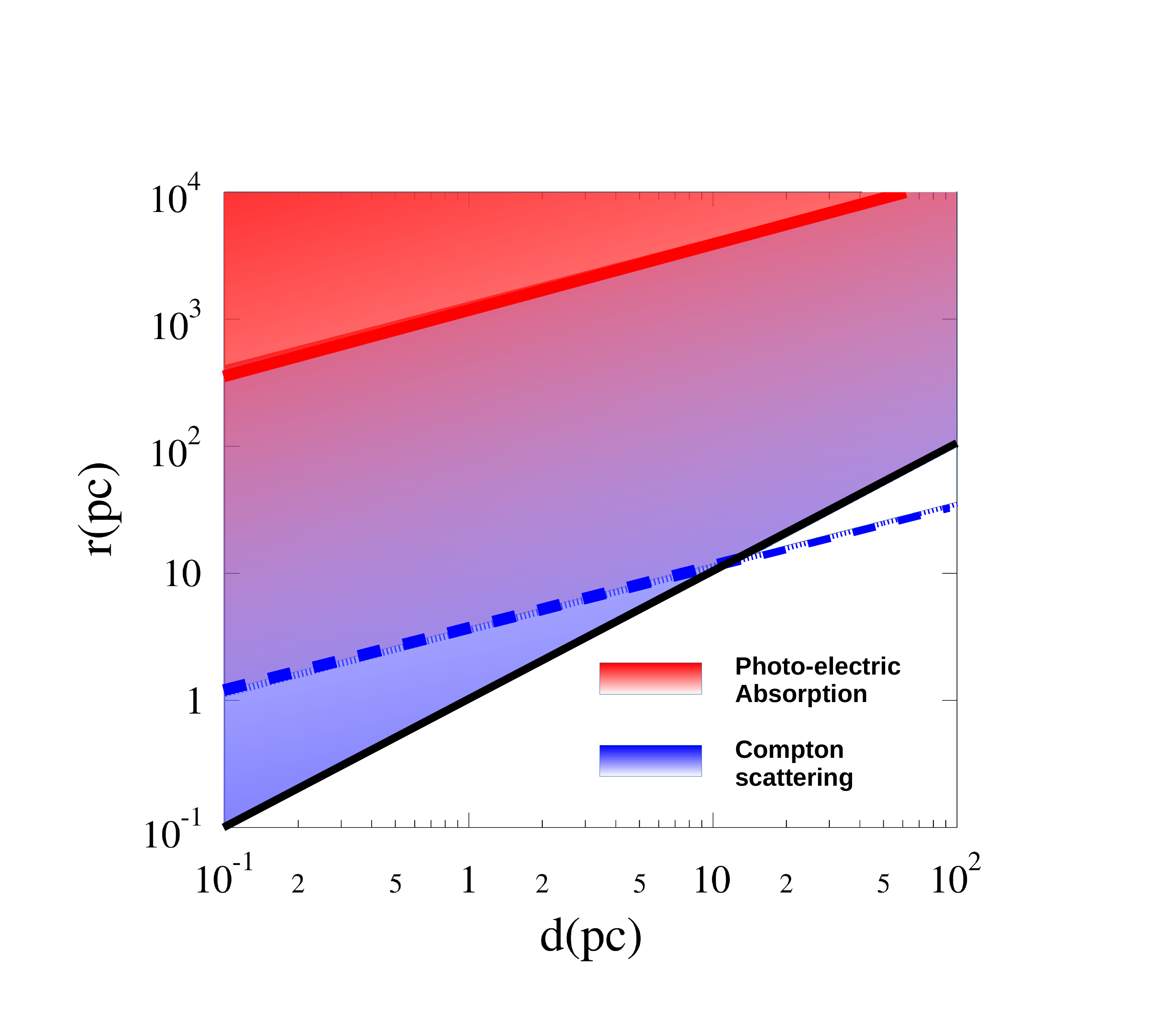}}
\caption{The main X-ray attenuation mechanisms as function of the distance $r$ from the AGN core to the dust cloud, and the thickness $d$ of the cloud. The (dashed) blue line indicates where a dust column equal to $X_{tot}=150\;\mathrm{g\;cm^{-2}}$, leading to a 90\% X-ray attenuation due to the Compton scattering process, would be highly ionized. The (full) red line indicates where a dust column equal to $X_{tot}=1.2\cdot 10^{-3}\;\mathrm{g\;cm^{-2}}$, corresponding to a 90\% X-ray attenuation due to the photo-electric absorption, would be highly ionized.}
\label{ionization_fig}
\end{figure}

For 1.24 keV X-rays the main attenuation process in neutral dust is due to photo-electric absorption. However, in case of dust being illuminated by the high luminosity AGN-jet, the dust will be ionized and consequently heated through the Compton scattering process. Hence, for highly ionized dust the main attenuation will be due to Compton scattering, which occurs with a much smaller cross-section compared to photo-electric absorption. Therefore, to determine the X-ray attenuation depth, we need to consider the amount of ionization for a dust cloud located at a distance $r$ from the AGN core~\cite{Rey95}. This can be estimated by the ionization parameter $U_X$ for an ionizing X-ray continuum. Typical values for this parameter range between $U_X=10^{-3}$ for a partly ionized dust cloud, up to $U_X > 0.1$ when the cloud is highly ionized. The ionization parameter is defined by~\cite{netzer2},
\begin{equation}
U_X=\int\limits_{E_1}^{E_2}\frac{L_E / E}{4\pi r^2 c\; n_N}dE,
\label{Ionpar}
\end{equation}
where $E_1$ and $E_2$ denote the energy limits of the ionizing continuum for the dust constituent, $N$, with nucleon number density $n_N$. The speed of light, $c$, is introduced to make $U$ dimensionless, and $L_E$ denotes the monochromatic luminosity per unit of energy. The $U$ parameter can be interpreted as the total number of ionizing photons per time unit which illuminates the dust or gas, divided by the gas particle number density, which gives an indication for the level of recombination. As an example, we consider the $U$ parameter as obtained from Quasar broad-line clouds~\cite{netzer2}. In this situation it can be shown that the number of ionized hydrogen atoms compared to neutral hydrogen scales like,
\begin{equation}
\frac{N_{H^+}}{N_H}\simeq 10^{5.3}U.
\end{equation}
The exact dust or gas composition close to an AGN is unknown, but likely contains elements ranging from Hydrogen to Iron giving energy limits equal to $E_1\approx0.1$~keV, $E_2\approx10$~keV. Ignoring the weak energy dependence of $L_E$ in this region, we can use Eq.~\ref{Ionpar} to determine the distance $r$ between the AGN-core and the dust for which the dust will be highly ionized ($U_X>0.1$),
\begin{equation}
r= \left( \frac{\log_e(E_2 / E_1) L_E}{4\pi c n_N U_X^2} \right)^{1/2}.
\end{equation}
The nucleon number density can be linked to the thickness of the dust cloud $d$ and the total column density $X_{tot}$ by,
\begin{equation}
n_N=\frac{X_{tot}N_{av}}{d <A_{mol}>},
\end{equation}
where $N_{av}$ is Avogadro's constant and $<A_{mol}>$ the average molecular weight. Using the given relations, we show in Fig.~\ref{ionization_fig} the main X-ray attenuation process given the distance $r$ between the AGN-core and the dust and the thickness of the dust cloud $d$. To illustrate the effect, in Fig.~\ref{ionization_fig} we consider a gas composition equal to that found in our atmosphere. Taking an atmospheric dust composition gives a convenient average over the expected dust or gas constituents ranging from Hydrogen to Iron in the environment of an AGN. Furthermore, for an atmospheric dust composition the proton interaction depth is known from the proton-air interaction depth determined by the Pierre Auger collaboration~\cite{auger_proton_air_xsec}. 

To produce Fig.~\ref{ionization_fig} we need the X-ray attenuation at 1.24 keV for both Compton scattering as well as the photo-electric absorption. These values are obtained from XCOM~\cite{XCOM}, and given by, $\lambda_X^{Compton}=65\;\mathrm{g\;cm^{-2}}$, and $\lambda_X^{PE}=5\cdot10^{-4}\;\mathrm{g\;cm^{-2}}$. The non-shaded region is given by the condition that the largest distance between the dust and the AGN can not be smaller than the thickness of the dust cloud itself, $r < d$. The (dashed) blue line indicates the distance at which a dust column equal to $X_{tot}=150\;\mathrm{g\;cm^{-2}}$, which would give rise to an X-ray attenuation of 90\% due to Compton scattering, becomes highly ionized ($U_x > 0.1$). The (full) red line indicates the distance at which a dust column equal to $X_{tot}=1.2\cdot10^{-3}\;\mathrm{g\;cm^{-2}}$, which would give rise to an X-ray attenuation of 90\% due to the photo-electric absorption, becomes highly ionized. 

Due to the high magnetic field in AGN environments (O($10^{4}$~Gauss)), cosmic-ray acceleration to ultra-high energies $E>10^{18}$~eV, can take place at (sub-)parsec scales~\cite{agnbook,hillas}. From Fig.~\ref{ionization_fig} it follows that for clouds close to a source the attenuation has to be due to Compton scattering. Such nearby, parsec scale, dust configurations have been considered in several studies~\cite{bianchi}. For a similar attenuation due to photoelectric absorption, a very thin, (sub)-parsec scale, dust or gas cloud located at kpc to Mpc distances would be needed, which implies an unlikely situation. Therefore, in this article we consider X-ray attenuation due to highly ionized parsec scale dust or gas clouds and in the following we use for the X-ray attenuation depth, $\lambda_X=\lambda_X^{Compton}$. 

Since the dust is expected to be fully ionized on parsec scales, one concern might be the stability of the dust configuration. Next to Compton heating, the dust might also be swept up by the relativistic jet or travel perpendicular to our line of sight, leaving the jet after a short time span. For example, dust clouds in Keplerian motion have been observed leading to strong X-ray variability on short time scales~\cite{bianchi_45}. Besides short lived configurations, in~\cite{bianchi} also more stable dust configurations over longer time scales are discussed. One interesting observation is that several studies have been performed to search for neutrino emission during flaring states where the X-ray or gamma-ray emission is strong, where in our model we expect the neutrino emission to be strongest during periods where the emission is weak in case the observed X-ray or gamma-ray attenuation is due to dust.  

So-far we considered a dust composition equal to the elements found in our atmosphere. The actual dust composition close to an AGN is badly known, but is expected to differ from the elements found in our atmosphere~\cite{mor_netzer}. It follows that we induce an uncertainty for both the X-ray attenuation as well as the proton interaction depth. To estimate the importance of this uncertainty, both quantities are determined considering different types of dust within the range between a pure Hydrogen and a pure Iron dust composition. Along with Hydrogen and Iron and the main atmospheric constituents Nitrogen and Oxygen, we also consider Silicon and Carbon which have been observed during infrared studies of AGN dust~\cite{mor_netzer}. The X-ray attenuation depths due to Compton scattering are obtained from the XCOM photon cross-section database. The proton-matter interaction depth is only known for air, but can be estimated for the different dust constituents. The proton interaction depth is linked to the proton-dust cross-section by, $\lambda_{p-N}=(\rho_N\sigma_{p-N})^{-1}$, where the amount of particles per unit mass is given by, $\rho_N\;\mathrm{[g^{-1}]}$, hence $\rho_N\propto A^{-1}$, where $A$ equals the average atomic mass number of the dust constituents. The proton-matter cross-section scales with the size of the nucleus given by $\sigma_{p-N}\propto A^{2/3}$. It follows that the proton interaction depth can be estimated to scale with atomic mass number as $\lambda_{p-N}=C(\mathrm{g\;cm^{-2}})A^{1/3}$. The scaling factor $C$ will be based on the proton-air interaction length obtained by the Pierre Auger collaboration and is found to be $C=23\;\mathrm{g\;cm^{-2}}$.
\begin{table}
\centering
\begin{tabular}{|c | c | c|c|}
\hline
Dust composition & $\lambda_{p-N}\;(\mathrm{g\;cm^{-2}}$) & $\lambda_X\;(\mathrm{g\;cm^{-2}}$) & $\lambda_{p-N} / \lambda_X$ \\
\hline
H(A=1)   & 23  & 14 &  1.6\\
\hline
C(A=12)  & 53 &  55 &  1.0\\
\hline
N(A=14)  & 56 &  62 &  0.9\\
\hline
O(A=16)  & 59 &  80 &  0.7\\
\hline
Si(A=28) & 71 &  55 &  1.3\\
\hline
Fe(A=56) & 89 &  84 &  1.1\\
\hline
\end{tabular}
\caption{The proton interaction depth $\lambda_{p-N}$ and the X-ray attenuation depth $\lambda_X$ for different dust compositions. The ratio between the proton interaction depth and the X-ray attenuation depth is also given.}
\label{table:dust_composition}
\end{table}

In Table~\ref{table:dust_composition}, the obtained values are shown. Also the ratio between the proton interaction depth and the X-ray attenuation depth is given. Taking an Nitrogen-Oxygen mixture as found for our atmosphere gives a ratio between the proton interaction depth and the X-ray attenuation depth due to Compton scattering in the range between $(\lambda_{p-air}/\lambda_X)=0.7-0.9$. The largest difference in ratio is with dust consisting out of pure Hydrogen, $(\lambda_{p-H}/\lambda_X)=1.6$ and the spread in values is small.

\section{Obscured flat spectrum radio AGN candidate selection}\label{blazar_selection}
In this section we specify a set of nearby obscured flat spectrum radio AGN. Under the assumption that the X-ray attenuation found for these objects is due to dust obscuration we investigate the additional high-energy neutrino production which is expected through the jet-matter interaction. The starting point of this investigation is based on two different galaxy catalogs, the Nijmegen catalog~\cite{nijmegen}, and the Fermi-2LAC catalog~\cite{fermi2lac}. First a red-shift selection is applied to assure an unbiased set of objects as outlined hereafter. Subsequently, a selection on the radio spectral index will be made to select a sample of flat spectrum radio AGN. In the following, as a first application of the method derived in the previous section, we select a set of objects based on their relative X-ray obscuration for which we determine the proton fraction which is expected to interact with the dust or gas surrounding the AGN. All data used in this section are retrieved from the NASA/IPAC Extragalactic Database (NED)~\cite{ned}.

\subsection{The Catalogs}
The starting point of this analysis is based on two source catalogs. The first catalog, the Nijmegen catalog~\cite{nijmegen}, contains 575 radio-galaxies and covers 88\% of the sky. The radio sources in this catalog have been separated into four different categories: Starforming Galaxies, Jets and Lobes, Unresolved Point Sources, and Unknown Morphology. The radio-flux measurements used to construct this catalog were taken from The NRAO VLA Sky Survey (NVSS)~\cite{nvss} observing at 1.4~GHz, and The Sydney University Molonglo Sky Survey (SUMSS)~\cite{sumss} observing at 843~MHz. NVSS covers the Northern sky above a declination of $-40^{\circ}$, while, SUMSS covers the Southern sky below a declination of $-30^{\circ}$. Both catalogs neglect galaxies within a galactic latitude of $|b| < 10^{\circ}$. The galaxy population was constructed to contain a volume-limited sample of strong radio sources which could be responsible for the UHECRs measured at Earth. The objects in this survey are located within several hundreds of Mpc, which means they are inside the close universe $z < 0.1$. Furthermore, an initial selection on the radio flux at 1.4~GHz ($>231$~mJy) and 843~MHz ($>289$~mJy) was made.

\begin{figure}
  \centering    
  \subfigure[The radio morphology of NGC 0262. A clear circular morphology is visible indicating that a possible jet is pointing toward us. This object falls into the Unresolved Point Source category of the Nijmegen catalog~\cite{nijmegen}.]{\label{fig:fig_sub1}\includegraphics[width=70mm]{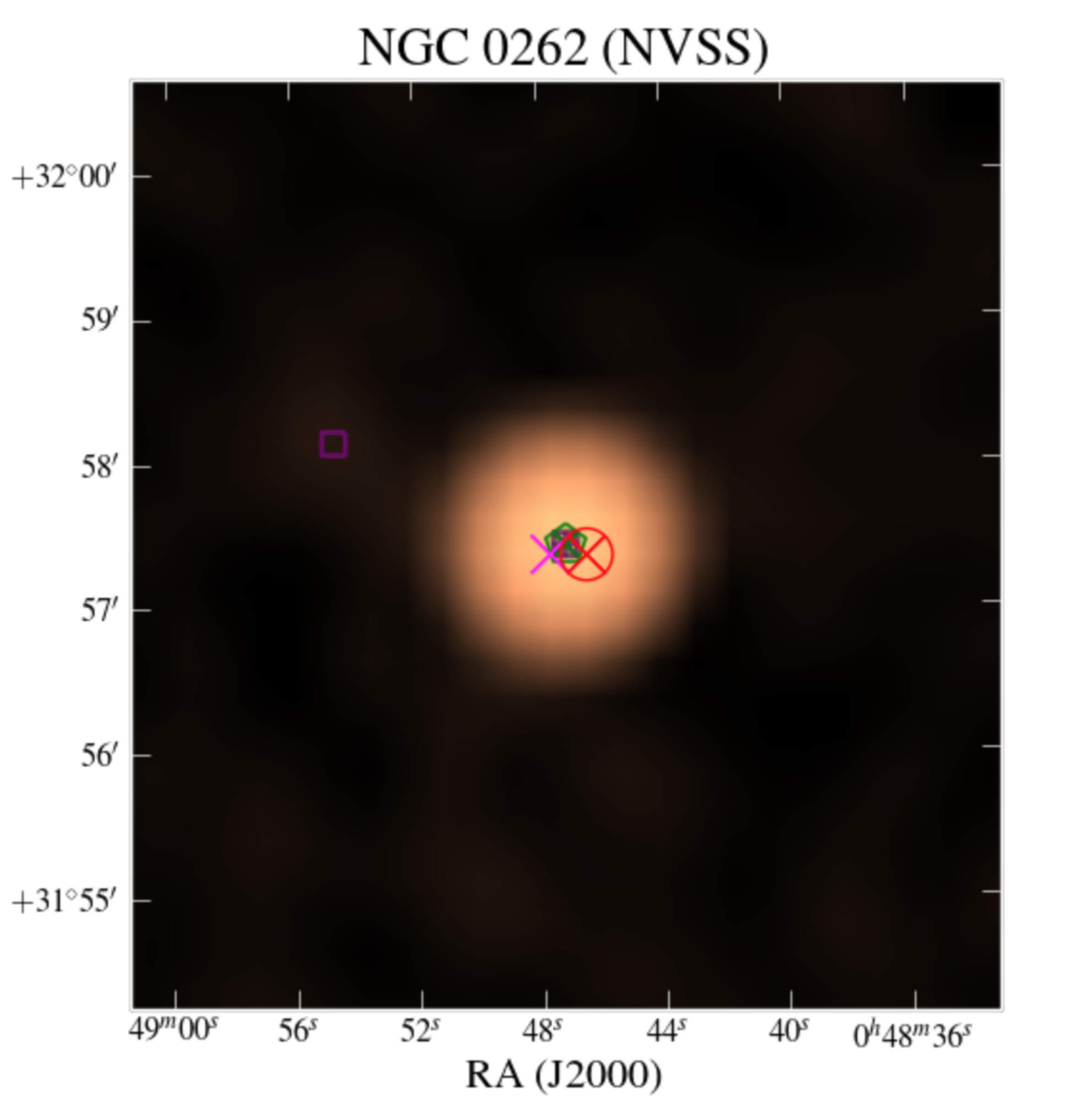}}
  \subfigure[The radio morphology of B2 0647+28. A clear jet structure is visible, indicating a non-Blazar object. This object falls into the Jets and Lobes category of the Nijmegen catalog~\cite{nijmegen}.]{\label{fig:fig_sub2}\includegraphics[width=70mm]{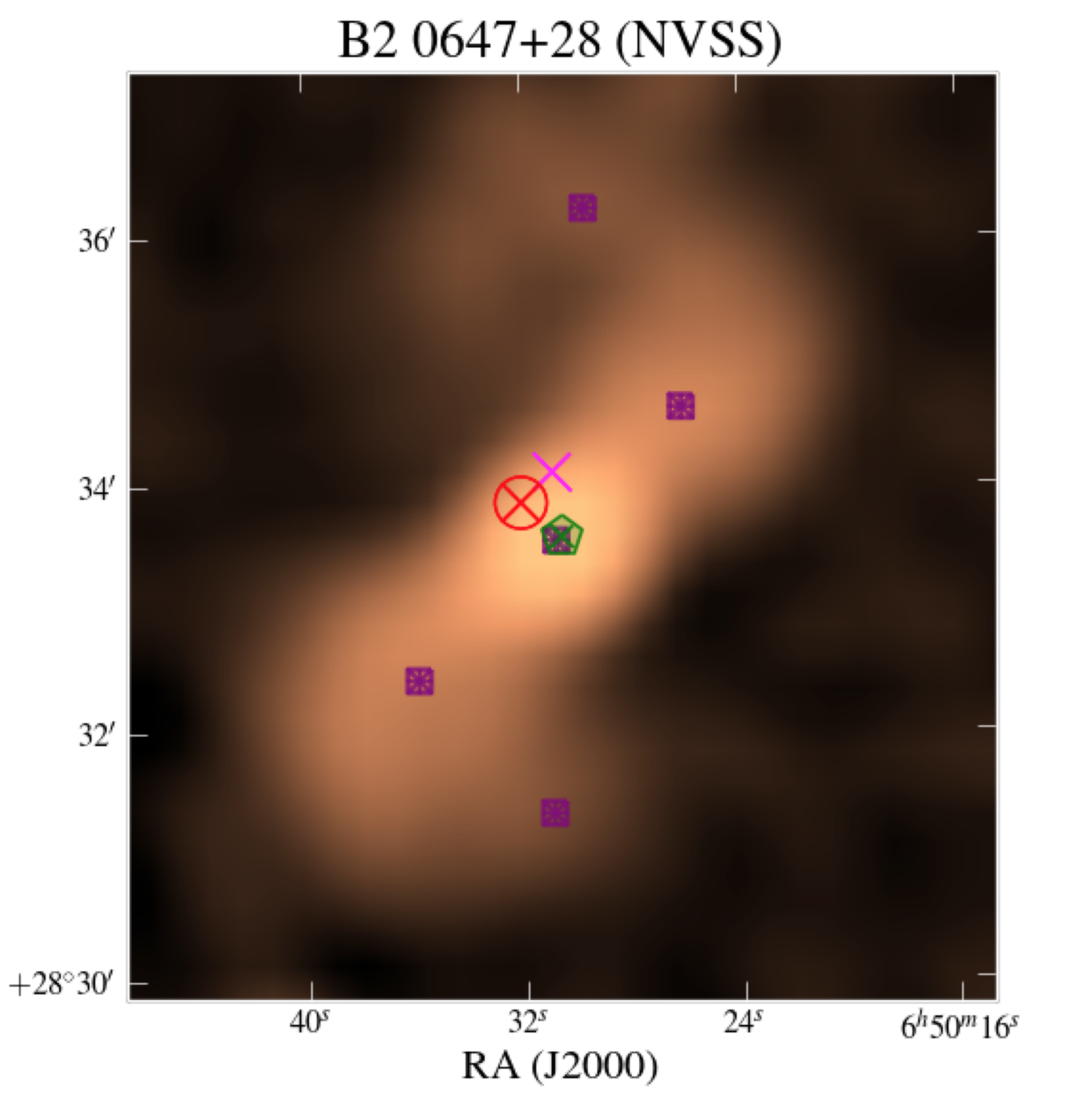}}
  \caption{The radio morphology for NGC 0262 and B2 0647+24. Data is obtained from NVSS~\cite{nvss}. Plots taken from~\cite{nijmegen}.}
\end{figure}

For the analysis presented in this paper we focus on the Starforming-Galaxy and Unresolved Point Source categories. The Jets and Lobes category, as well as the Unknown Morphology category by definition indicate objects of which the jet does not point in our direction, and are therefore excluded. An example is shown in Fig.~\ref{fig:fig_sub1} which shows an object from the unresolved point-source category. The circular radio-morphology indicates a possible jet pointing toward us. In Fig.~\ref{fig:fig_sub2} we observe one of the objects in the jets and lobes category, which has a clear non-circular radio morphology indicating a jet not pointing toward us.

The second source sample is based on The Second Catalog of Active Galactic Nuclei Detected by the Fermi Large Area Telescope (2LAC) \cite{fermi2lac}, which is formed by 1017 $\gamma$-ray sources. The 2LAC catalog is divided in the following categories: Active Galactic Nuclei, Active Galaxies of Uncertain type, BL Lacertae objects, Flat-Spectrum Radio Quasars, Radio Galaxies, Steep-Spectrum Radio Quasars, Starburst Galaxies and Unidentified sources. The 2LAC sample includes objects with a galactic latitude $|b| > 10^{\circ}$, and with a redshift $z<3.1$. From the 2LAC sample only objects with red-shift information in NED were selected.  

\subsection{Composing an unbiased set of objects}
After cross-correlating the objects in both catalogs, an initial sample of 735 unique galaxies remains. The main goal of our selection is to specify a set of potentially obscured flat spectrum radio AGN. We will specify these objects based on the relative X-ray attenuation with respect to the emission at lower (radio) frequencies. To assure that the X-ray attenuation is due to a possible dust component and not due to attenuation by the intergalactic medium, redshift effects, or a possible selection effect, we determine in this section the red-shift for which an unbiased set of objects is detected in the X-ray band. This selection will be performed for the objects in the 2LAC catalog, which contains objects with a redshift $z<3.1$, whereas the Nijmegen catalog is already limited to the close universe $z<0.1$. 

\begin{figure}[htb]
  \begin{minipage}[b]{\linewidth}
    \centering\includegraphics[width=0.9\textwidth]{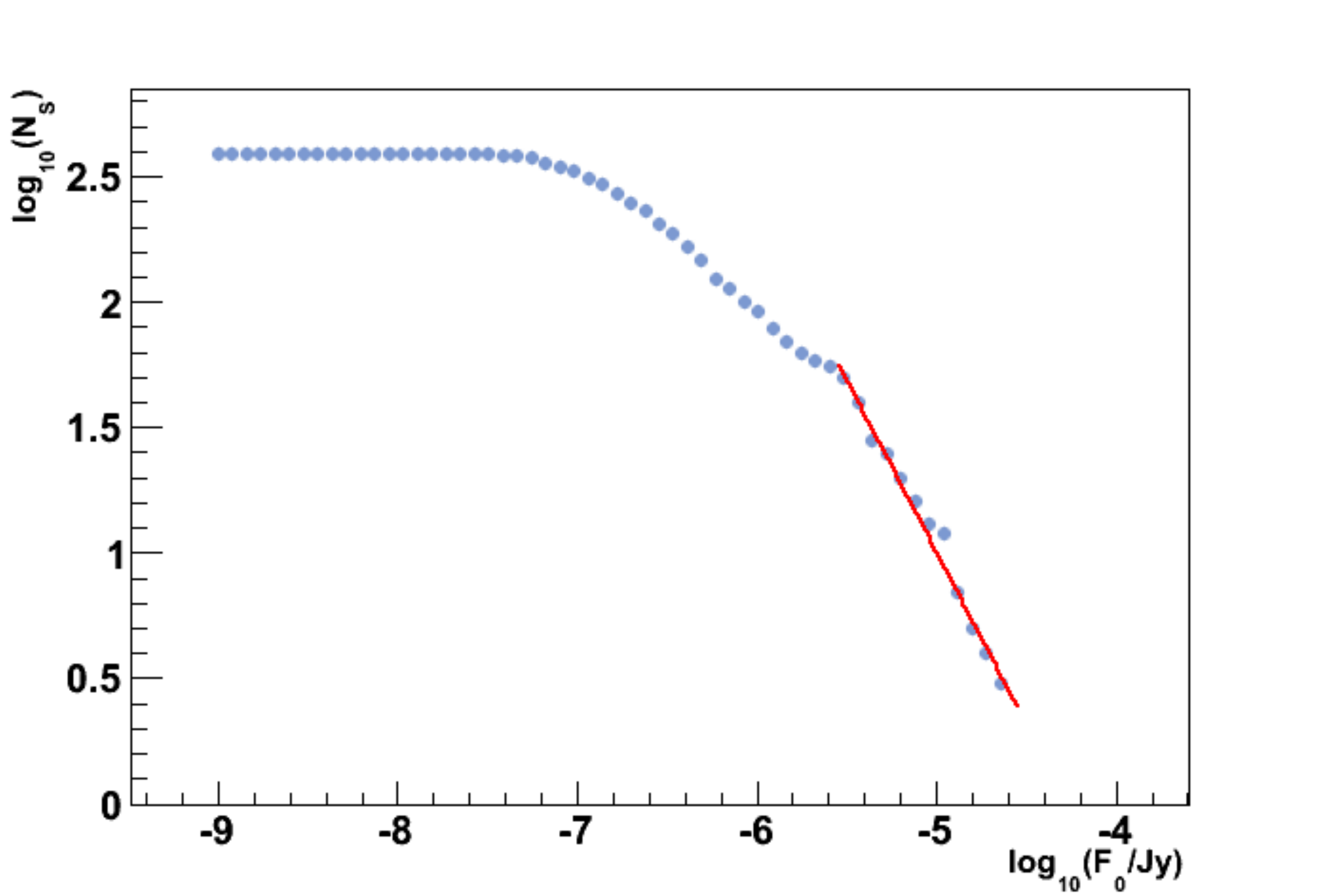}
    \caption{The number of sources $N(F_m > F_0)$, with a flux $F_m$ larger than $F_0$ at $\nu=3.02$$\times$$10^{17}$~Hz corresponding to the 1.24 keV X-ray band. The full (red) line shows the fit with a slope equal to $-1.36\pm 0.06$ compatible to -3/2 within two standard deviations as expected for an unbiased set of objects.}
    \label{cumulative_nsources}
  \end{minipage}
\end{figure}

In case of a uniform source density and a generic luminosity we expect the number of sources with a measured flux $F_m$ above a threshold value $F_0$, to scale like,
\begin{equation}
N(F_m > F_0) \propto (F_0)^{-3/2}.
\end{equation}
Fig.~\ref{cumulative_nsources} shows the number of sources with a flux $F_m > F_0$. The flux is determined for the 1.24~keV X-ray band corresponding to a frequency of $3.02$$\times$$10^{17}$~Hz. The measurements are obtained from~\cite{ned}. Indeed, one observes the expected power-law behavior between $-5.5 \leq \log_{10}(F_{0}/\mathrm{Jy}) \leq -4.6$. At smaller threshold values the distribution flattens indicating a deficit of sources.
\begin{figure}[htb]
  \begin{minipage}[b]{\linewidth}
    \centering\includegraphics[width=0.9\textwidth]{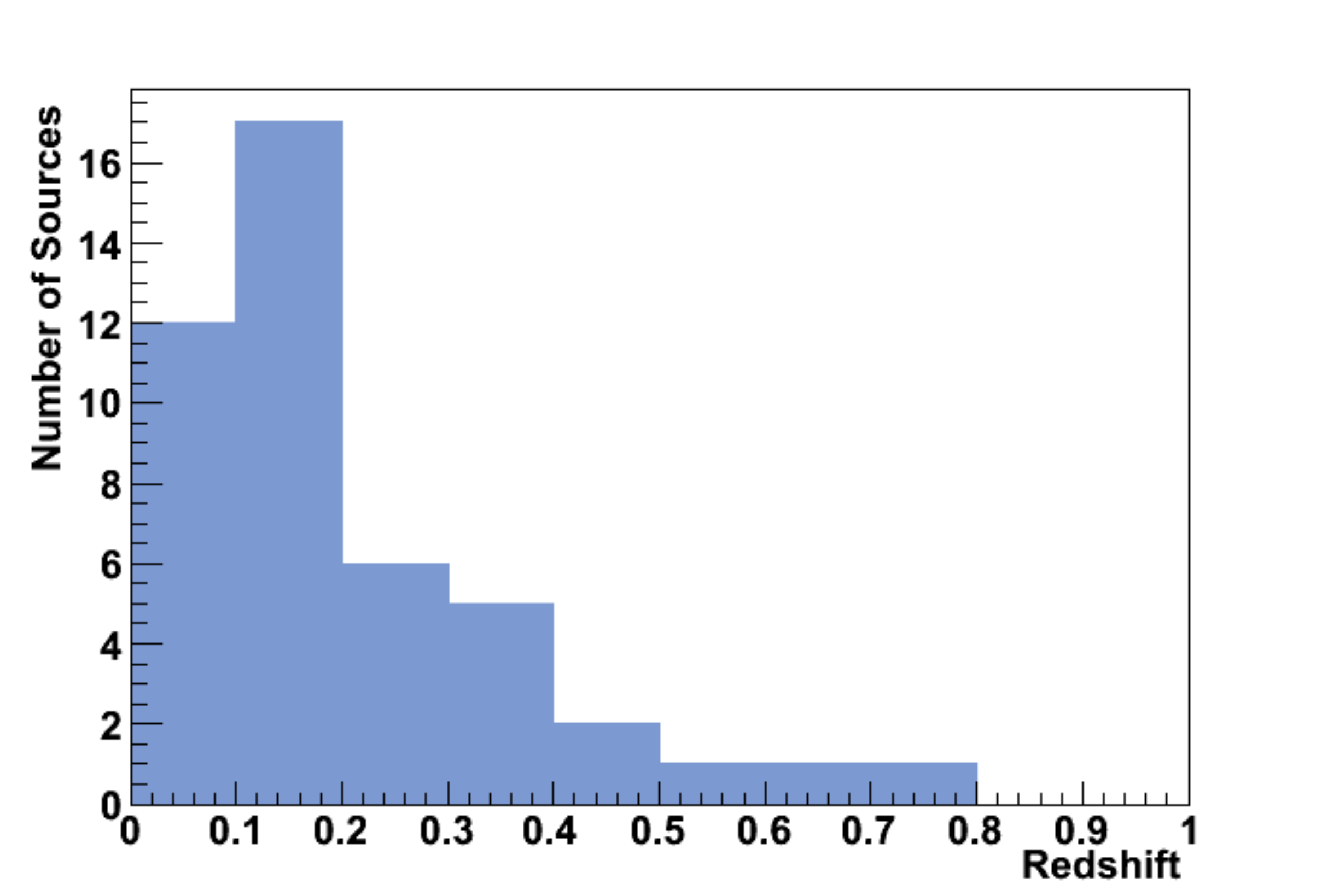}
    \caption{The redshift distribution of the sources with an observed flux larger than $\log_{10}(F_{m}/\mathrm{Jy}) \geq -5.5$.}
    \label{redshift_cut}
  \end{minipage}
\end{figure}
To determine the redshift for which an unbiased set of objects is expected, we plot the redshift distribution of the sources with a measured flux larger than $\log_{10}(F_{m}/\mathrm{Jy}) \geq -5.5$ in Fig.~\ref{redshift_cut}. To assure an unbiased set of objects, in the following we select all objects (without imposing any condition on their flux) with a redshift smaller than the median of this distribution ($z<0.17$), after which 209 objects remain. 

\subsection{Flat spectrum radio AGN selection}
A Blazar is defined as an AGN with its relativistic jet pointing toward us. For such an object typically strong, variable, emission up to high frequencies is expected. Nevertheless, in case the central engine is blocked by surrounding matter, the high-energy electromagnetic emission will be suppressed. Therefore, we base our selection on the observed radio spectrum. A typical signature for an AGN with its jet pointing toward us is the relativistic boosting of the synchrotron emission to high frequencies. Experimentally this is probed by the radio spectral index, $\alpha_{R}$, at frequencies of a few GHz, for which the measured flux is expected to follow a power law given by $F_\nu=C\nu^{\alpha_R}$~Jy~\cite{agnbook}. Objects with a frequency spectral index $\alpha_{R}>-0.5$, are typically classified as Flat Spectrum Radio Quasars, or BL Lac objects~\cite{agnbook}.
      
After applying the redshift selection we remain with 209 objects. For these objects we apply a flux-fit within the frequency range $\nu=0.843-5$~GHz on the available data found in NED~\cite{ned}. The typical uncertainty on the obtained data points lies between 1-20\%. Therefore, in case no information about the uncertainty of the measurement is given by NED, a rather conservative value of 20\% is assumed. Since in general a high-power AGN is expected to have variable emission over different time scales, the obtained fit might be affected by data taken at different periods in time. To obtain consistent results, we perform an initial fit after which we iteratively exclude data points which deviate more than 4$\sigma$ from the fit, and we re-fit the remaining data.
\begin{figure}
  \centering    
  \subfigure[The flux-fit for ARP 220. The radio spectral index is given by $\alpha_R=-0.2415\pm0.01662$, indicating a possible AGN-jet is pointing toward us.]{\includegraphics[width=70mm]{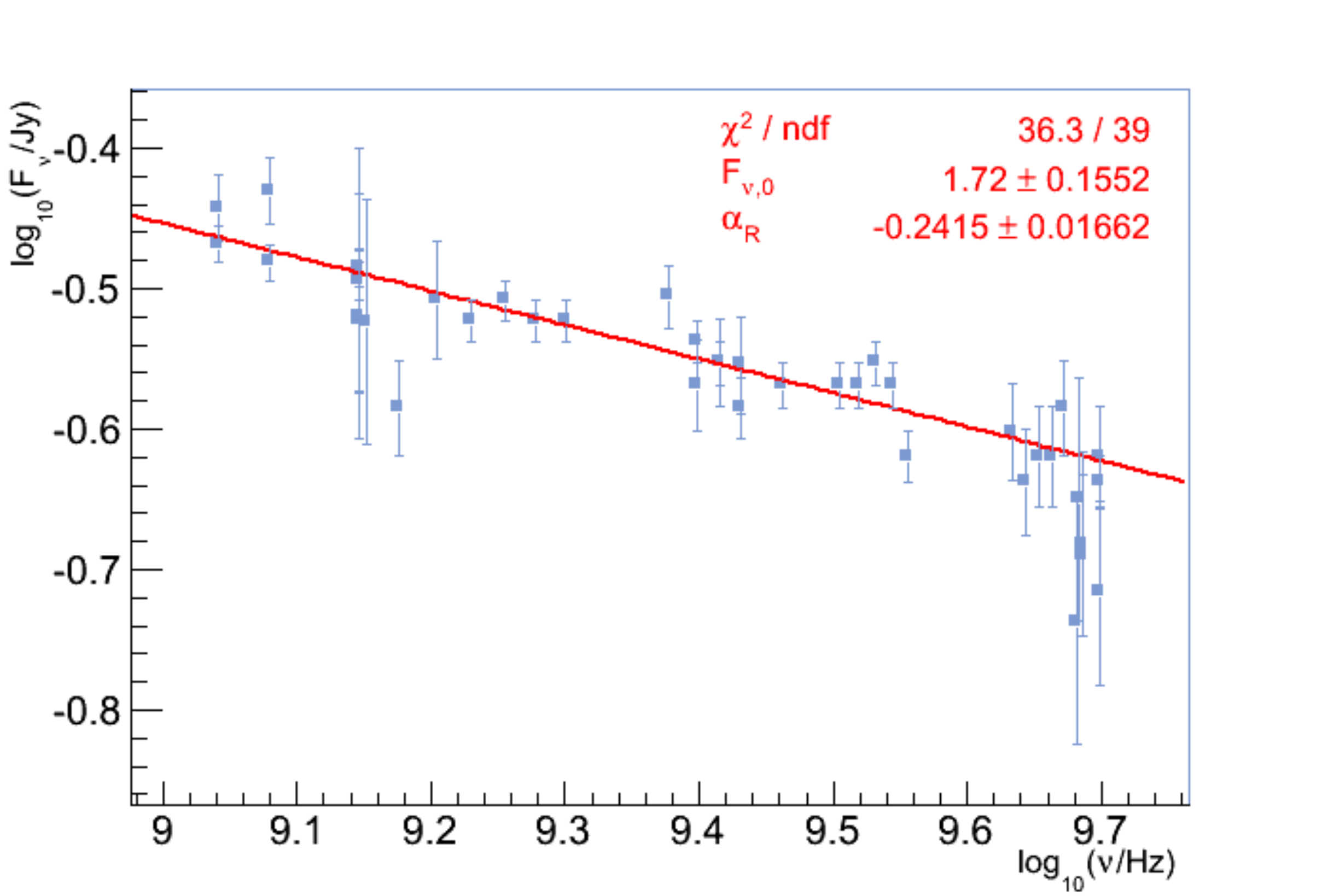}}
  \subfigure[The flux-fit for NGC 7674. The radio spectral index is given by $\alpha_R=-0.8801\pm0.05799$. The steep radio spectrum indicates that very likely there is no jet directed toward us.]{\includegraphics[width=70mm]{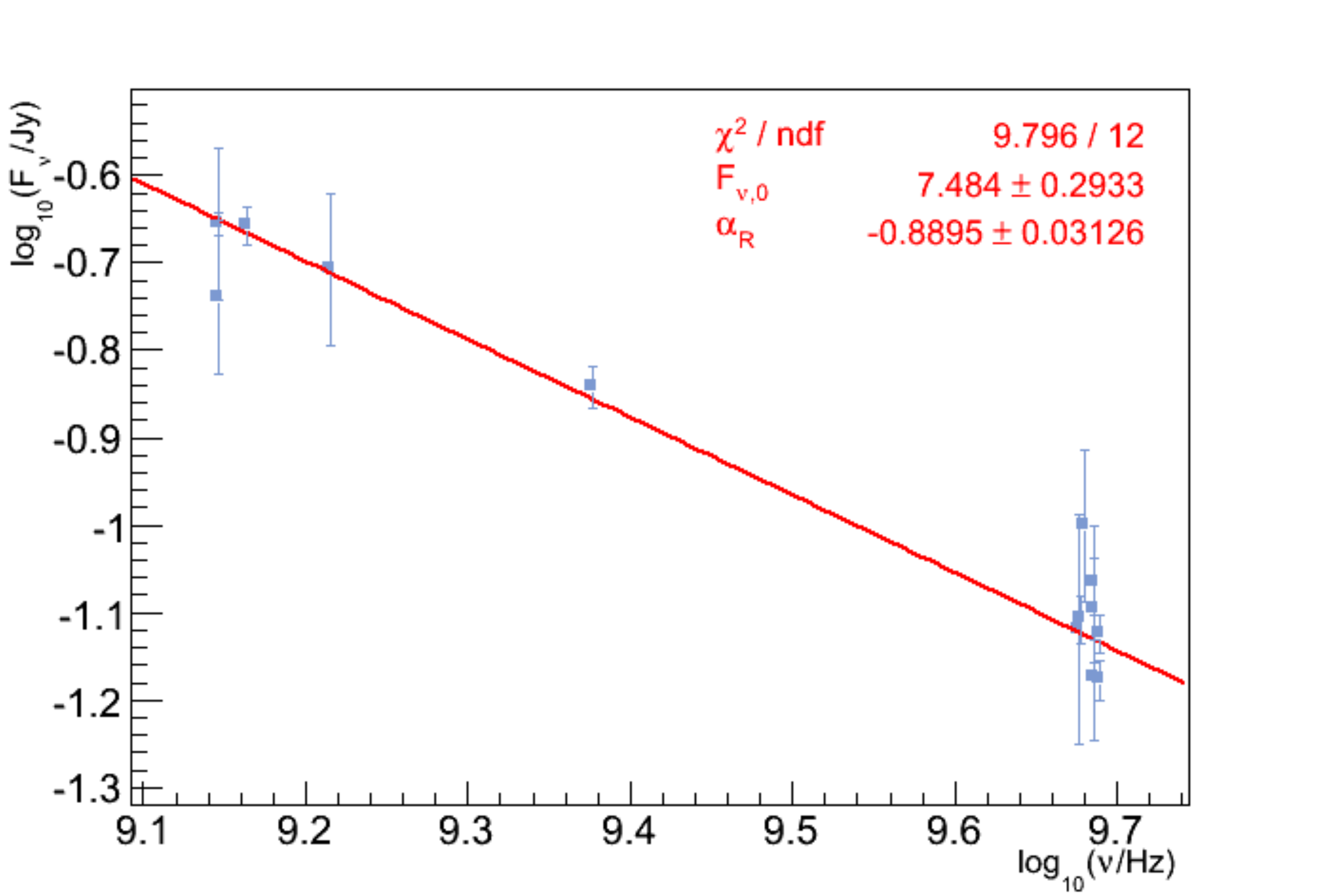}}
  \caption{The flux-fit for ARP 220 (a) and NGC 7674 (b).}
  \label{fits:both}
\end{figure}

\begin{figure}[htb]
  \begin{minipage}[b]{\linewidth}
    \centering\includegraphics[width=0.9\textwidth]{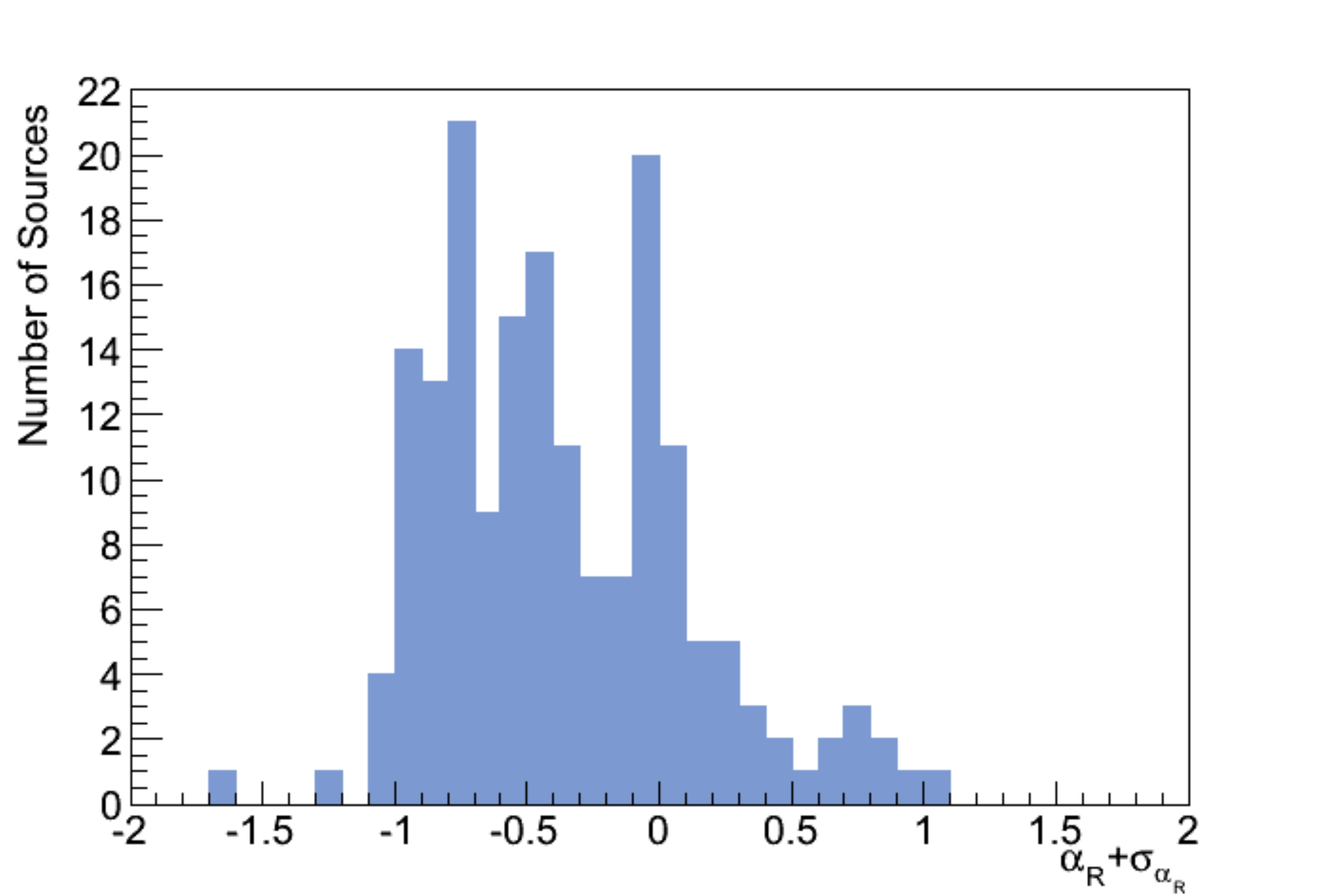}
    \caption{Distribution of the frequency spectral index $\alpha_{R}+\sigma_{\alpha_R}$.}
    \label{alpha}
  \end{minipage}
\end{figure}

Fig.~\ref{fits:both} shows two examples of the fit performance, where ARP 220 is accepted, and NGC 7674 fails to pass the radio spectral index selection. After the removal of objects with a lack of data, we select sources with a radio spectral index $\alpha_{R}+\sigma_{\alpha_{R}} > -0.5$. A distribution of the obtained spectral indices is shown in Fig.~\ref{alpha}. It follows that a large number of galaxies fail to pass our selection, and hence are excluded from our analysis. After this selection, we end up with a set of 98 nearby sources which can be classified as flat spectrum radio AGN.     

\subsection{Source classes}
In case of AGN obscuration by surrounding dust or gas, emission at low (radio) frequencies is expected to be unattenuated, whereas emission at higher frequencies in the X-ray and $\gamma$-ray bands will be attenuated. Therefore, to determine the amount of obscuring matter, we will focus on the measured X-ray luminosity relative to the radio luminosity. This will be done considering the monochromatic luminosity at $1.4$~GHz for the radio band and at $3.02$$\times$$10^{17}$~Hz corresponding to the 1.24~keV X-ray band. To X-ray measurements for which no error is given in~\cite{ned} an error of 17\% is attributed, equal to the average error found in the 1.24 keV X-ray band. These bands have been selected on basis of the abundance of measurements at these frequencies. Nevertheless, for most objects in the Southern sky there are no measurements at $\nu=1.4$~GHz since the data for these objects are obtained from the SUMSS survey which operates at $\nu=843$~MHz. For these objects, the flux value at $\nu=1.4$~GHz is obtained from the radio-flux fit presented previously.

After removal of objects with a lack of data at 1.24~keV, a sample of 62 flat spectrum radio AGN remains, of which 49 are located in the Northern sky and 13 are located in the Southern sky. The 62 remaining flat spectrum radio AGN can be sub-divided in roughly three categories: Flat Spectrum Radio Quasars (FSRQs) (14 objects), Ultra Luminous Infrared Galaxies (ULIRGs) (3 objects), and BLLac objects (45 objects). These objects are given in Appendix A.

\begin{figure}[htb]
  \begin{minipage}[b]{\linewidth}
    \centering\includegraphics[width=0.9\textwidth]{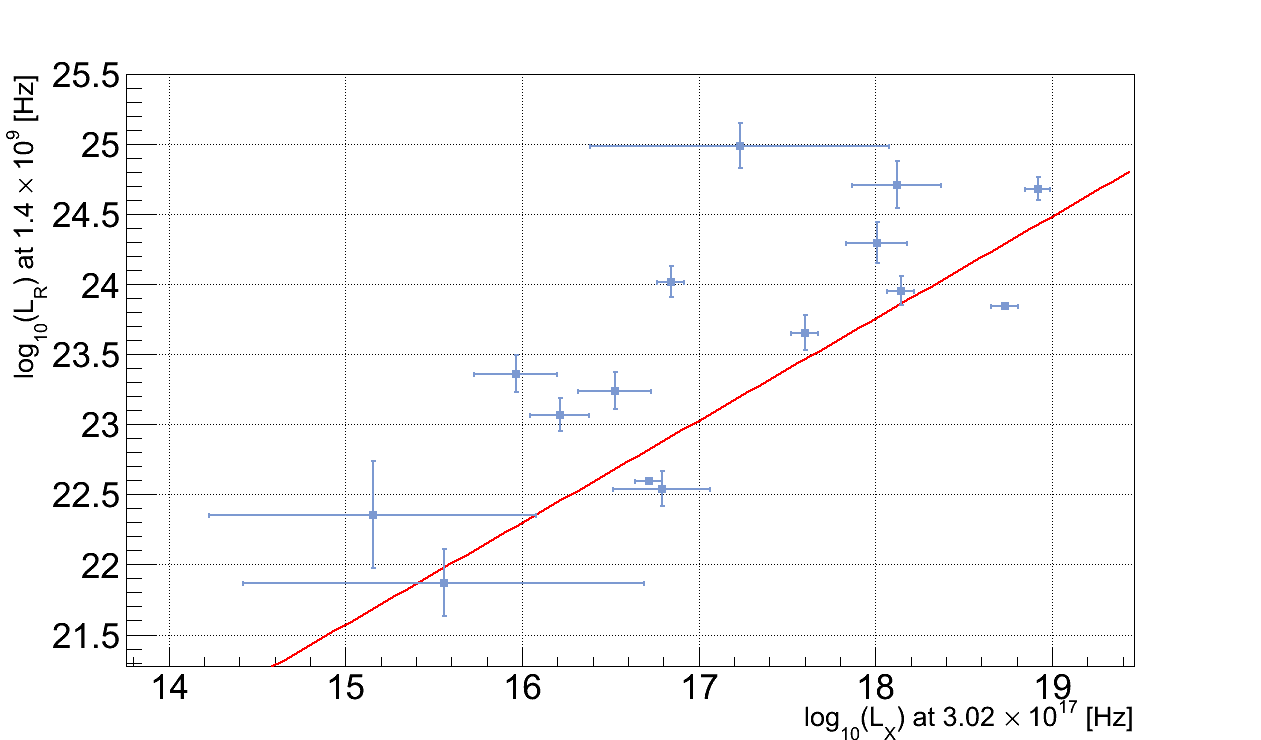}
    \caption{The observed radio luminosity as function of the observed X-ray luminosity for the FSRQ and ULIRG objects. The full (red) line gives the fit to this distribution with slope value $\beta=0.73\pm0.04$}
    \label{r_vs_x_fsrq}
  \end{minipage}
\end{figure}
\begin{figure}[htb]
  \begin{minipage}[b]{\linewidth}
    \centering\includegraphics[width=0.9\textwidth]{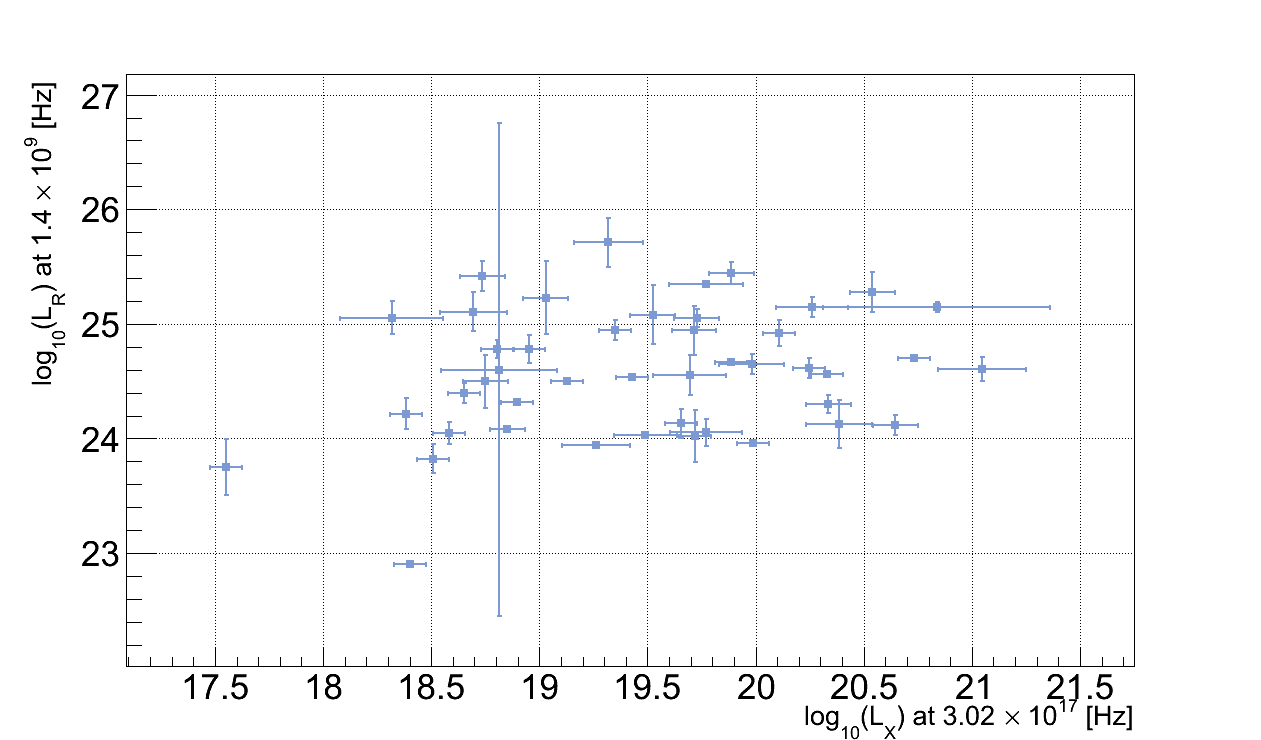}
    \caption{The observed radio luminosity as function of the observed X-ray luminosity for the BLLac objects. No clear correlation between the radio luminosity and the X-ray luminosity is present.}
    \label{r_vs_x_bllac}
  \end{minipage}
\end{figure}

Most of the objects remaining from the Nijmegen catalog can be classified as Flat Spectrum Radio Quasars. The SED of these objects is determined by a typical two-bump structure, where the first bump is given by synchrotron emission from the high-energy AGN-jet. The second bump is shifted to higher frequencies. The emission in the X-ray band for this source class is linked to the emission from the hotter parts of the accretion disc. For this source class a clear relation between the radio luminosity $L_R$ and the X-ray luminosity $L_X$ is given by a power law relation $L_{R}=L_{X}^{\beta}$, with the power $\beta$ ranging between $\beta = 0.6 - 0.7$~\cite{falcke2003,mer2003}. In~\figref{r_vs_x_fsrq}, we plot the radio luminosity against the X-ray luminosity for the FSRQ as well as the ULIRG objects in our selection. From the power law fit we obtain $\beta = 0.73\pm0.04$, which clearly reflects the expected behavior. Since we are searching for X-ray attenuation due to surrounding matter, we have to correct for this relation and in the following we will consider the intensity ratio $I_{F-U}=L_X^{0.73}/L_R$ for these object classes. One special object is 3C273, which is very bright in both radio as well as X-ray. There are strong indications that the X-ray emission for this object is dominated by the jet and not due to the accretion disc~\cite{Uch06}. It follows that the radio luminosity correction should not be made for this object. Due to its very bright X-ray emission, the object is by definition not of interest for our selection and will therefore not be considered.

The ULIRG objects are very interesting on their own. Next to the standard two-bump spectrum, these objects have a very bright emission in the infrared-UV range of the spectrum. These objects (ARP 220, NGC 3628, NGC 3690), with ARP 220 as the most prominent ULIRG, are interacting galaxies for which a large amount of dust is observed, indicated by the very bright infrared-UV emission. Besides the infrared-UV peak, the radio emission as well as the emission in the X-ray and gamma-ray bands for our selected objects show similar behavior as found for the FSRQ objects. NGC 3628 is a well known object, which has been studied in detail. The X-ray measurements given in~\cite{ned} for this object are within errors compatible with background, which makes the data not suitable for our analysis. It should be noted however, that since the X-ray measurements for this object are compatible with background a very strong dust component is expected to block the high-energy emission which makes this object very interesting to study on its own.
The ULIRG object class, and especially ARP 220, is still under detailed investigation~\cite{williams2010,wilson2014}, where indications have been found that an AGN is present in, at least some of these objects~\cite{wilson2014}. Even though it is not determined whether or not there is an AGN located inside ARP 220 and NGC 3690, we decided to keep them in our selection based on the fact that strong (flat spectrum) radio emission has been observed which is most likely due to synchrotron emission. Hence particle acceleration takes place at these sources. This in combination with the very strong dust component, makes them very interesting candidates for high-energy neutrino production through the proton-matter interaction channel.

BLLac objects are distinguished from FSRQs by a featureless flat to rising spectrum up to very high frequencies. The emission from these objects is believed to be jet-dominated, for which the relation $L_{R}=L_{X}^{\beta}$ is less clear~\cite{falcke2003,mer2003}. This is also observed in~\figref{r_vs_x_bllac}, where we plot the radio luminosity $L_R$ as a function of the X-ray luminosity $L_X$ for the remaining BLLac objects. Even though in~\cite{falcke2003} a correction factor linked to the relativistic boosting is given, from~\figref{r_vs_x_bllac} no clear correlation between the radio and X-ray emission is observed. Therefore, contrary to the FSRQ and ULIRG objects, in the following we will base our selection on the X-ray emission without correcting relative to the radio luminosity for BLLac objects. 
\begin{table*}[Ht]
{\renewcommand\arraystretch{1.25}
\begin{tabular}{|l|l|l|l|l|l|l|} \hline
 \textbf{Object name (NED ID)}   & \textbf{H} & \textbf{C} & \textbf{N} & \textbf{O} & \textbf{Si} & \textbf{Fe} \\ \hline
 \textbf{Class: FSRQ}&&&&&&\\ \hline
2MASXJ05581173+5328180&0.79&0.93&0.94&0.97&0.87&0.91 \\ \hline
CGCG186-048&0.71&0.88&0.90&0.94&0.79&0.85 \\ \hline
MRK0668&0.92&0.99&0.99&1.00&0.96&0.98 \\ \hline
\textbf{Class: ULIRG}&&&&&&\\ \hline
ARP220&0.79&0.93&0.94&0.97&0.86&0.91 \\ \hline
\textbf{Class: BLLac}&&&&&&\\ \hline
3C371&0.84&0.95&0.96&0.98&0.90&0.94 \\ \hline
B21811+31&0.81&0.94&0.95&0.98&0.88&0.92 \\ \hline
SBS0812+578&0.87&0.97&0.98&0.99&0.93&0.96 \\ \hline
GB6J1542+6129&0.82&0.95&0.96&0.98&0.89&0.93 \\ \hline
RGBJ1534+372&0.86&0.97&0.97&0.99&0.92&0.95 \\ \hline
SBS1200+608&0.89&0.98&0.98&0.99&0.94&0.97 \\ \hline
PKS1349-439&0.85&0.96&0.97&0.98&0.91&0.95 \\ \hline
4C+04.77&0.97&1.00&1.00&1.00&0.99&1.00 \\ \hline
1H1720+117&0.89&0.98&0.98&0.99&0.94&0.97 \\ \hline
APLibrae&0.90&0.98&0.99&0.99&0.95&0.97 \\ \hline
PKS1717+177&0.83&0.95&0.96&0.98&0.89&0.93 \\ \hline

\end{tabular}}
\caption{The fraction of protons which is expected to interact for the 25\% weakest X-ray objects (indicated by the full (red) lines in Fig.~\ref{r_vs_x_fsrq_rat} and Fig.~\ref{r_vs_x_bllac_rat}) for the different source categories. This fraction is given for different dust constituents following the estimated proton-dust interaction depth given in Table~\ref{table:dust_composition}.}
\label{table:obscured_table}
\end{table*}

\subsection{Source candidate investigation}\label{dust_obscured}
\begin{figure}[htb]
  \begin{minipage}[b]{\linewidth}
    \centering\includegraphics[width=0.9\textwidth]{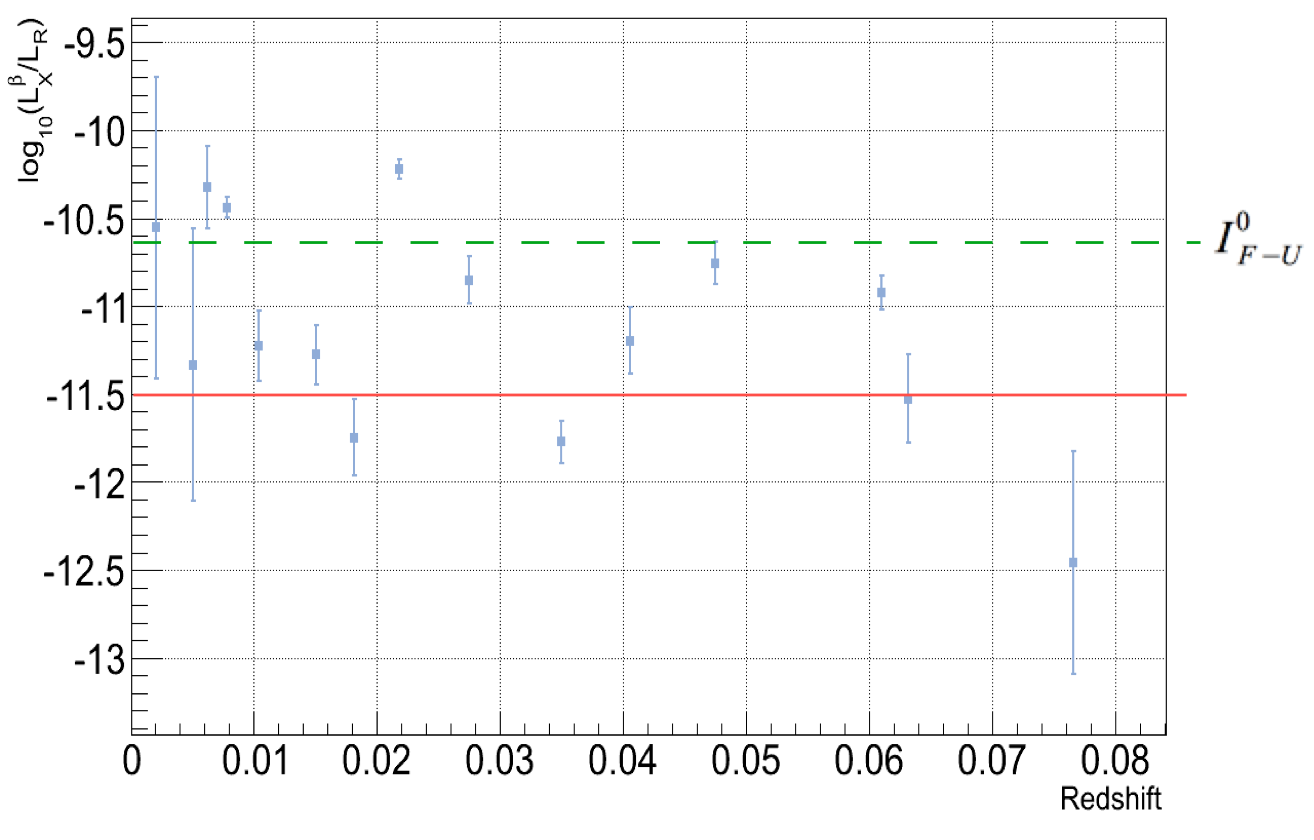}
    \caption{The corrected X-ray intensity ratio $I_{F-U}=L_X^{0.73}/L_R$ as a function of redshift for the FSRQ and ULIRG objects. The base value $I^0_{F-U}$ defined by the 25\% strongest X-ray sources is given by the dashed (green) line. The 25\% weakest X-ray sources are located below the full (red) line.}
    \label{r_vs_x_fsrq_rat}
  \end{minipage}
\end{figure}
\begin{figure}[htb]
  \begin{minipage}[b]{\linewidth}
    \centering\includegraphics[width=0.9\textwidth]{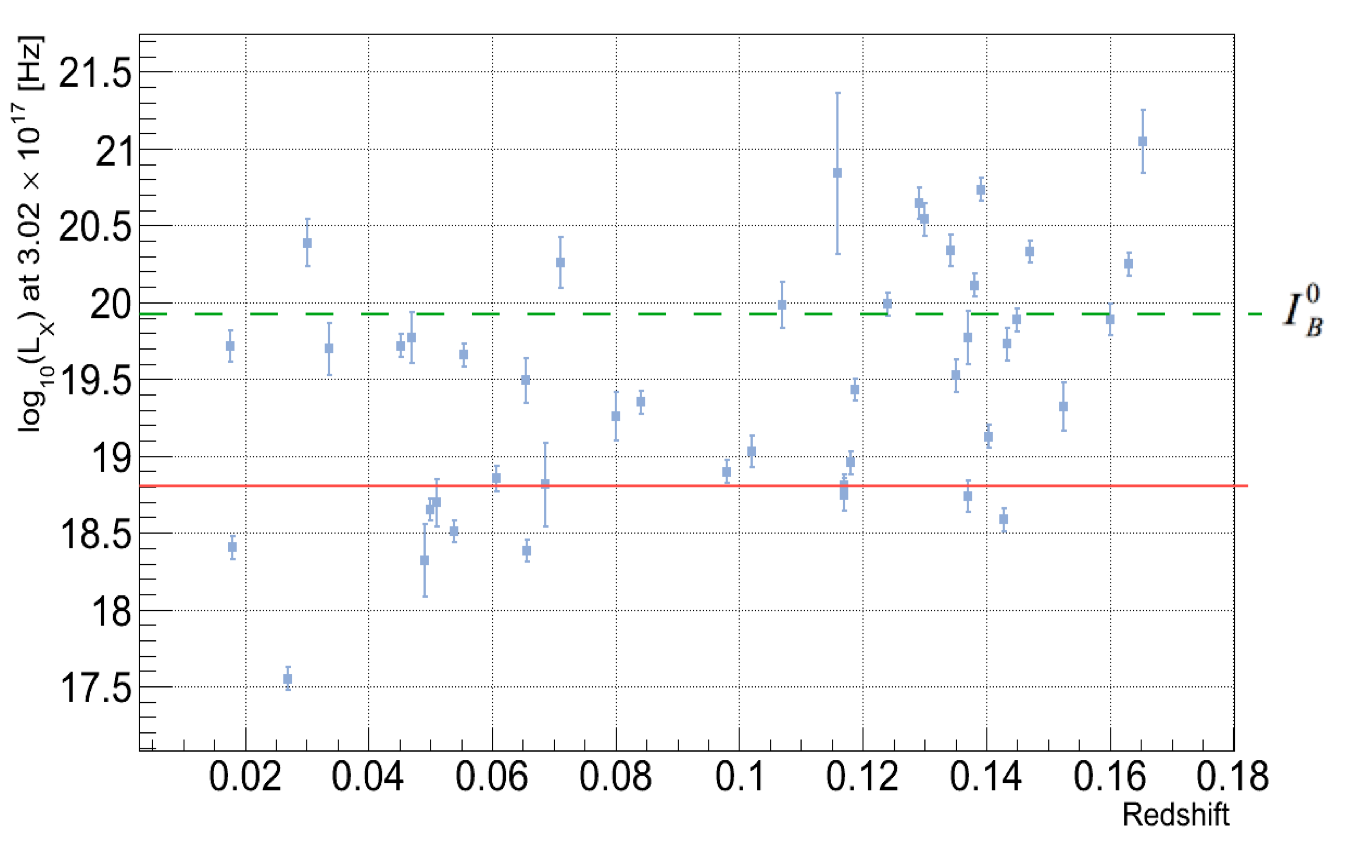}
    \caption{The X-ray intensity $I_{B}=L_X$ as a function of redshift for the BLLac objects. The base value $I^0_{B}$ defined by the 25\% strongest X-ray sources is given by the dashed (green) line. The 25\% weakest X-ray sources are located below the full (red) line.}
    \label{r_vs_x_bllac_rat}
  \end{minipage}
\end{figure}
The amount of protons interacting with the dust or gas will depend on the total column density $X_{tot}$. In section~\ref{proton_matter_subsec} a method was developed to determine this column density based on the observed X-ray intensity $I_X^{obs}$ relative to a base value $I_X^0$ for non-obscured AGN. Furthermore, it was shown that the X-ray attenuation for surrounding matter on (sub-)parsec scales will be due to Compton scattering for which the attenuation length $\lambda_X^{Compton}$ is given in Table~\ref{table:dust_composition} for different dust constituents.

This leaves us with determining a base value $I_X^{0}$. In the determination of this base value we separate our sample, based on the relation between the radio luminosity and the X-ray luminosity as discussed above. Hence, we consider FSRQs and ULIRGs on one side and BLLac objects on the other side. For the first category we consider the corrected intensity ratio, $I_{F-U}=L_X^{0.73}/L_R$, where for the BLLac objects we simply consider the X-ray luminosity $I_{B}=L_X$.

To investigate the possibility of additional neutrino production we assume that the observed spread in (relative) X-ray intensity is fully attributed to surrounding dust. Under this assumption, the base value for $I^0$ is now determined by the 25\% strongest X-ray objects which are considered to be unobscured. The obtained base values are given by the intensity ratio $I^0_{F-U}=2.3\cdot 10^{-11}$ indicated by the dashed (green) line in Fig.~\ref{r_vs_x_fsrq_rat} for the FSRQ and ULIRG sample, and the intensity $I^0_B=9.7\cdot10^{19}$ indicated by the dashed (green) line in Fig.~\ref{r_vs_x_bllac_rat} for the BLLac sample.

Having defined our base value $I^0_X$ we are able to calculate the fraction of protons which is expected to interact with the dust or gas from the measured X-ray intensity $I_X^{obs}$. Following the derivation in section~\ref{proton_matter_subsec}, and combining Eq.~\ref{Ip} and Eq.~\ref{Ix} we find,
\begin{equation}
\frac{I_p^{int}}{I^0_p}=1-\left(\frac{I_X^{obs}}{I^0_{X}}\right)^{\lambda_X / \lambda_{p-N}}.
\end{equation}
In Table~II, we show the fraction of protons which is expected to interact for the 25\% weakest X-ray objects (located below the full lines in Fig.~\ref{r_vs_x_fsrq_rat} and Fig.~\ref{r_vs_x_bllac_rat}) for the different source categories. This fraction is given for different dust constituents following the estimated proton-dust interaction depth given in Table~\ref{table:dust_composition}.

We observe that in case the observed X-ray attenuation is fully due to surrounding dust, we indeed expect to have a strong increase in neutrino production from these objects, which makes them very interesting to investigate as possible cosmic neutrino sources. 

\section{Summary}
In this article we discuss a new sub-class of possible high-energy neutrino sources. This sub-class is composed of AGN with their high-energy jet pointing toward Earth, which is obscured by dust or gas surrounding the central engine. We refer to these objects as obscured flat spectrum radio AGN. In case a hadronic component is accelerated inside the high-energy AGN-jet, the jet-matter interaction will lead to additional neutrino production. We argue that this production channel is independent of the photo-hadronic neutrino production given by a hadronic component in the high-energy jet interacting with the ambient photon flux. Furthermore, it is shown that in a full beam-dump scenario which occurs for surrounding matter with a column density equal to several proton interaction depths, the neutrino production increases up to an order of magnitude. 

One specific property of this sub-class is that the neutrino emission mainly occurs through the jet-matter interaction. It follows that due to the damping of the hadronic component in the jet, we expect an anti-correlation between the cosmic neutrinos and ultra-high-energy cosmic rays detected at Earth. Next to the hadronic component, also the high-energy X-ray and gamma-ray emission is expected to be attenuated for this source class.  

A method has been constructed to select possible obscured flat spectrum radio AGN by determining the column density of the surrounding matter. The method is based on the attenuation of emission at small wavelengths in the X-ray band, where emission at long wavelengths in the radio band is expected to pass through the dust or gas unattenuated.

As a first application of the devised method, two different source catalogs are investigated to specify a set of possible obscured flat spectrum radio AGN. The first catalog, the Nijmegen-catalog, contains a volume limited sample of strong radio galaxies. The second catalog that is investigated is the Fermi-2LAC catalog. Starting point of the selection procedure is a red-shift cut to assure an unbiased set of objects. Subsequently, the radio spectral index of the remaining objects was determined
to assure the high-energy AGN jet was pointing toward us. After these initial cuts three different source classes can be identified, Flat Spectrum Radio Quasars (FSRQs), Ultra Luminous Infra-Red Galaxies (ULIRGs) on one side and BLLac objects on the other side. Our sample was split between the FSRQ and ULIRG objects, and the BLLac objects based on their different relation between radio luminosity and X-ray luminosity.

Finally, we investigate the possibility for these objects to have an increased neutrino production. Under the assumption that the observed X-ray attenuation is due to surrounding dust or gas, we find that for the 25\% weakest X-ray sources, 50-100\% of the surviving hadrons in the jet are expected to interact with the surrounding dust or gas. This would give rise to a significant increase in neutrino production, which makes these objects very interesting to investigate as possible sources for the detected high-energy cosmic neutrino flux.

\acknowledgements
The authors would like to acknowledge the following funding agencies for their support of the research presented in this report: The Flemish Foundation for Scientific Research (FWO-12L3715N - K.D. de Vries), the  FWO  Odysseus  program (G.0917.09. - N. van Eijndhoven), the 'aspirant FWO Vlaanderen' and Strategic Research Program ‘High Energy Physics’ of the Vrije Universiteit Brussel (M. Vereecken), the European Research Council (ERC) under the European Union’s Horizon 2020 research and innovation programme (grant agreement No 640130 - S. Buitink). This research has made use of the NASA/IPAC Extragalactic Database (NED) which is operated by the Jet Propulsion Laboratory, California Institute of Technology, under contract with the National Aeronautics and Space Administration.


\cleardoublepage

\onecolumngrid
\appendix
\section{Flat spectrum radio AGN}\label{appendix:blazar_feature}
The shown values are:  declination (dec), right ascension (ra), redshift (z), frequency spectral index ($\alpha_R+\sigma_{\alpha_R}$) between 843~MHz and 5~GHz, the radio flux $f_{\nu}$(Jy) at $1.4$$\times$$10^{9}$ [Hz] (radio) and the X-ray flux at $3.02$$\times$$10^{17}$(Jy) [Hz] (X-ray) for the 62 selected flat spectrum radio AGN.
\begin{table*}[Hb]
{\renewcommand\arraystretch{1.00}
\begin{tabular}{|l|l|l|l|l|l|l|} \hline
    \textbf{Object name (NED ID)}   &   \textbf{dec}    &   \textbf{ra}     &   \textbf{z}\boldmath$\times10^2$     &   \boldmath$\alpha_R+\sigma_{\alpha_R}$  & \textbf{$\mathbf{f_\nu^{radio}}$[Jy]}  &   \textbf{$\mathbf{f_\nu^{X-ray}}$[Jy]} \\
    \hline
    1ES1215+303&30.12&184.47&13.00&-0.06&0.45&8.25e-06 \\ \hline
NGC5506&-3.21&213.31&0.62&-0.44&0.34&6.15e-07 \\ \hline
PKS2155-304&-30.23&329.72&11.60&-0.08&0.44&2.18e-05 \\ \hline
3C371&69.82&271.71&5.10&0.08&2.26&8.79e-07 \\ \hline
B21811+31&31.74&273.40&11.70&0.05&0.19&1.96e-07 \\ \hline
RBS0958&20.24&169.28&13.92&0.98&0.10&1.11e-05 \\ \hline
NGC1275&41.51&49.95&1.76&0.51&14.60&8.65e-05 \\ \hline
SBS0812+578&57.65&124.09&5.39&-0.36&0.10&5.02e-07 \\ \hline
Mkn180&70.16&174.11&4.53&-0.20&0.23&1.16e-05 \\ \hline
PKS0447-439&-43.84&72.35&10.70&0.42&0.16&3.54e-06 \\ \hline
PMNJ0152+0146&1.79&28.17&8.00&-0.05&0.06&1.28e-06 \\ \hline
1H0323+022&2.42&51.56&14.70&1.04&0.07&3.99e-06 \\ \hline
NGC2110&-7.46&88.05&0.78&-0.40&0.30&3.97e-07 \\ \hline
NGC1052&-8.26&40.27&0.50&0.86&0.59&3.78e-08 \\ \hline
GB6J1542+6129&61.50&235.74&11.70&0.31&0.10&1.71e-07 \\ \hline
TXS2320+343&34.60&350.68&9.80&-0.09&0.10&3.61e-07 \\ \hline
GB6J1053+4930&49.50&163.43&14.04&0.69&0.06&2.72e-07 \\ \hline
3C273&2.05&187.28&15.83&0.04&42.61&1.88e-05 \\ \hline
1H0323+342&34.18&51.17&6.10&-0.37&0.59&1.03e-06 \\ \hline
B32247+381&38.41&342.52&11.87&0.72&0.10&8.11e-07 \\ \hline
NGC4278&29.28&185.03&0.21&-0.19&0.40&1.97e-07 \\ \hline
TXS1148+592&58.99&177.85&11.80&-0.09&0.18&2.67e-07 \\ \hline
OQ530&54.39&214.94&15.26&0.26&0.87&3.54e-07 \\ \hline
ARP220&23.50&233.74&1.81&-0.22&0.32&1.30e-08 \\ \hline
Mkn421&38.21&166.11&3.00&-0.01&0.66&0.000 \\ \hline
NGC3690&58.56&172.13&1.04&-0.49&0.71&1.37e-07 \\ \hline
RGBJ1534+372&37.27&233.70&14.28&0.25&0.02&7.55e-08 \\ \hline
1ES1440+122&12.01&220.70&16.31&0.24&0.06&2.59e-06 \\ \hline
BLLacertae&42.28&330.68&6.86&-0.04&0.39&6.38e-07 \\ \hline
SBS1200+608&60.52&180.76&6.56&-0.10&0.17&2.49e-07 \\ \hline
2MASXJ05581173+5328180&53.47&89.55&3.50&-0.38&0.40&2.65e-08 \\ \hline
WComae&28.23&185.38&10.20&-0.40&0.68&4.30e-07 \\ \hline
ON246&25.30&187.56&13.50&0.34&0.27&7.35e-07 \\ \hline
1H1914-194&-19.36&289.44&13.70&-0.06&0.48&1.28e-06 \\ \hline
PKS1349-439&-44.21&208.24&5.00&0.10&0.44&7.94e-07 \\ \hline
S31741+19&19.59&265.99&8.40&-0.40&0.55&1.40e-06 \\ \hline
PKS2005-489&-48.83&302.36&7.10&-0.08&1.26&1.63e-05 \\ \hline
1ES0806+524&52.32&122.45&13.80&0.01&0.18&2.73e-06 \\ \hline
PKS1424+240&23.80&216.75&16.00&-0.27&0.43&1.18e-06 \\ \hline

\end{tabular}}
\label{table:obscured_table}
\end{table*}
\begin{table*}[Hb]
{\renewcommand\arraystretch{1.00}
\begin{tabular}{|l|l|l|l|l|l|l|} \hline
    \textbf{Object name (NED ID)}   &   \textbf{dec}    &   \textbf{ra}     &   \textbf{z}\boldmath$\times10^2$     &   \boldmath$\alpha_R+\sigma_{\alpha_R}$   &   \textbf{$\mathbf{f_\nu^{radio}}$[Jy]}  &   \textbf{$\mathbf{f_\nu^{X-ray}}$[Jy]} \\
    \hline
    4C+04.77&4.67&331.07&2.70&-0.27&0.40&2.54e-07 \\ \hline
RXJ1136.5+6737&67.62&174.13&13.42&0.05&0.04&4.88e-06 \\ \hline
NGC0262&31.96&12.20&1.50&0.77&0.29&4.08e-08 \\ \hline
H1426+428&42.67&217.14&12.91&-0.13&0.03&1.08e-05 \\ \hline
TXS1055+567&56.47&164.66&14.33&0.03&0.22&1.04e-06 \\ \hline
Mkn501&39.76&253.47&3.37&-0.12&1.48&2.05e-05 \\ \hline
1H1720+117&11.87&261.27&1.80&0.74&0.12&3.80e-06 \\ \hline
RBS0970&42.20&170.20&12.40&0.25&0.02&2.58e-06 \\ \hline
NGC6521&62.61&268.95&2.75&-0.24&0.29&2.58e-07 \\ \hline
IZw187&50.22&262.08&5.54&-0.25&0.20&6.75e-06 \\ \hline
UGC03927&59.68&114.38&4.05&-0.24&0.55&2.86e-07 \\ \hline
IC4374&-27.02&211.87&2.18&-0.11&0.65&5.06e-06 \\ \hline
NGC3628&13.59&170.07&0.28&-0.30&0.34&1.74e-09 \\ \hline
1ES1959+650&65.15&300.00&4.70&-0.004&0.24&1.26e-05 \\ \hline
PMNJ0847-2337&-23.62&131.76&6.07&0.08&0.14&8.48e-07 \\ \hline
APLibrae&-24.37&229.42&4.90&-0.02&2.10&3.88e-07 \\ \hline
MG1J010908+1816&18.27&17.28&14.50&0.60&0.09&1.49e-06 \\ \hline
H2356-309&-30.63&359.78&16.54&-0.33&0.06&1.61e-05 \\ \hline
CGCG186-048&35.02&176.84&6.31&-0.08&0.56&1.46e-07 \\ \hline
PKS1717+177&17.75&259.80&13.70&-0.03&0.57&1.18e-07 \\ \hline
MRK0668&28.45&211.75&7.66&0.86&0.72&1.27e-08 \\ \hline
SBS1646+499&49.83&251.90&4.75&-0.01&0.18&2.88e-07 \\ \hline
PKS1440-389&-39.14&220.99&6.55&0.07&0.11&3.16e-06 \\ \hline

\end{tabular}}
\label{table:obscured_table}
\end{table*}

\end{document}